\documentclass[useAMS,usenatbib,usegraphicx,fleqn]{mn2e}			\usepackage{times}
\usepackage{natbib} 
\usepackage[]{graphicx,color}
\usepackage{amsmath}
\usepackage{amssymb}
\usepackage{textcomp}
\usepackage[normalem]{ulem}
\usepackage{german}
\usepackage{rotating}

\title[Bulge--Disc Decomposition of IFU Datacubes]
{SDSS-IV MaNGA: Bulge--Disc Decomposition of IFU Datacubes (BUDDI)}
\author[E. J. Johnston et al]{Evelyn~J.~Johnston$^{1,2}$\thanks{Email: evelyn.johnston@eso.org}, Boris H\"au\ss ler$^{1,3,4}$,  Alfonso~Arag\'on-Salamanca$^2$,  
\newauthor Michael~R.~Merrif\mbox{}ield$^2$, Steven Bamford$^2$, Matthew A. Bershady$^{5}$, Kevin Bundy$^{6}$, 
\newauthor Niv Drory$^7$, Hai~Fu$^8$, David Law$^9$, Christian Nitschelm$^{10}$, Daniel Thomas$^{11}$, 
\newauthor  Alexandre~Roman Lopes$^{12}$, David Wake$^{5,13}$, Renbin Yan$^{14}$\\
  $^1$European Southern Observatory, Alonso de C\'ordova 3107, Casilla 19001, Santiago, Chile\\ 
  $^2$School of Physics and Astronomy, University of Nottingham, University Park, Nottingham, NG7 2RD, UK\\
  $^3$University of Oxford, Denys Wilkinson Building, Keble Road, Oxford, Oxon OX1 3RH, UK\\
  $^4$University of Hertfordshire, Hatfield, Hertfordshire AL10 9AB, UK\\
  $^{5}$Department of Astronomy, University of Winsconsin-Madison, 475 N. Charter Street, Madison, WI 53706-1582, USA.\\
  $^{6}$Kavli Institute for the Physics and Mathematics of the Universe (WPI), The University of Tokyo Institutes for Advanced Study, The University of Tokyo, \\Kashiwa, Chiba 277-8583, Japan\\
  $^7$McDonald Observatory, University of Texas at Austin, Austin, TX 78712, USA\\
  $^8$Department of Physics \& Astronomy, University of Iowa, Iowa City, IA 52245\\
  $^9$Space Telescope Science Institute, 3700 San Martin Drive, Baltimore, MD 21218, USA\\
  $^{10}$Unidad de Astronom\'ia, Universidad de Antofagasta, Avenida Angamos 601, Antofagasta 1270300, Chile\\
  $^{11}$Institute of Cosmology and Gravitation, University of Portsmouth, Portsmouth, UK\\
  $^{12}$Departamento de F\'isica, Facultad de Ciencias, Universidad de La Serena, Cisternas 1200, La Serena, Chile\\
  $^{13}$Department of Physical Sciences, The Open University, Milton Keynes, MK7 6AA, UK.\\
  $^{14}$ Department of Physics and Astronomy, University of Kentucky, 505 Rose Street, Lexington, KY 40506, USA.\\
}

\begin{document}

\maketitle

\begin{abstract}
With the availability of large integral-field unit (IFU) spectral surveys of nearby galaxies, there is now the potential to extract spectral information from across the bulges and discs of galaxies in a systematic way. This information can address questions such as how these components built up with time, how galaxies evolve and whether their evolution depends on other properties of the galaxy such as its mass or environment. We present \textsc{buddi}, a new approach to fit the two-dimensional light profiles of galaxies as a function of wavelength to extract the spectral properties of these galaxies' discs and bulges. The fitting is carried out using \textsc{GalfitM}, a modified form of \textsc{Galfit} which can fit multi-waveband images simultaneously. The benefit of this technique over traditional multi-waveband fits is that the stellar populations of each component can be constrained using knowledge over the whole image and spectrum available. The decomposition has been developed using commissioning data from the SDSS-IV Mapping Nearby Galaxies at APO (MaNGA) survey with redshifts z~$<$0.14 and coverage of at least 1.5~effective radii for a spatial resolution of 2.5~arcsec FWHM and field of view of $>$22~arcsec, but can be applied to any IFU data of a nearby galaxy with similar or better spatial resolution and coverage. We present an overview of the fitting process, the results from our tests, and we finish with example stellar population analyses of early-type galaxies from the MaNGA survey to give an indication of the scientific potential of applying bulge--disc decomposition to IFU data.

\end{abstract}

\begin{keywords}
  galaxies: structure -- 
  galaxies: evolution --
  galaxies: formation -- 
  galaxies: bulges --
  galaxies: stellar content --
\end{keywords}

\section{Introduction}\label{sec:introduction}

The assembly history of a galaxy can be studied through its stellar populations - the spectral features provide information about the star formation history of the galaxy, while the spatial distribution of the different stellar populations gives clues as to how they have been redistributed throughout the galaxy during mergers and secular processes. 
Components such as the bulge and disc formed through different mechanisms, and thus have experienced fundamentally different star-formation histories \citep{Seigar_1998,Macarthur_2003, Allen_2006, Cameron_2009, Simard_2011}. Consequently, many studies have tried to separate the light from the bulge and disc to better understand their formation and contribution to the evolution of the host galaxy. 

One first step towards a complete decomposition of a galaxy is its separation into its two main components, the bulge and disc. So-called bulge--disc decomposition can be carried out in two ways- through one-dimensional surface brightness profile fitting \citep{Kormendy_1977, Burstein_1979, Kent_1985}, in which the radially-averaged light profile is fitted after correcting for the inclination of the galaxy, or two-dimensional fitting of images of galaxies \citep{Shaw_1989, Byun_1995, deJong_1996}. Both techniques provide information on the colours, sizes and luminosities of each component included in the fit, which provide constraints on the masses, ages and metallicities of each component \citep{Pforr_2012}. For example, as part of the fundamental plane, it has been found that in elliptical galaxies and bulges of disc galaxies the scale lengths increase and the effective surface brightnesses decrease with increasing luminosity \citep{Kormendy_1977, Hamabe_1987, 	
Djorgovski_1987}. These results suggest that both structures have experienced similar star-formation histories, despite the differing presence or absence of a dynamically cold disc in the host galaxies. However, while the same trend has been detected in the bulges of field S0s, it is not always present in cluster S0s, thus indicating that the local environment plays a significant role in the formation of S0 bulges \citep{Barway_2009}.

It has also been found that as you move from S0s to late-type spirals, the effective size and surface brightness of the bulge decrease, whereas discs in these galaxies display much less variation in these properties between morphologies \citep{deJong_1996_2,Graham_2001,Mollenhoff_2001,Trujillo_2002,Aguerri_2004, Laurikainen_2007,Graham_2008,Oohama_2009}. As a result, it is becoming apparent that the increase in the bulge-to-total light ratio observed in S0s is driven by the evolution of the bulge as opposed to simply the fading of the disc.

In recent years, bulge--disc decomposition has been increasingly applied to multi-waveband photometry since the bulge and disc colours can act as a proxy for their stellar populations. Comparisons of  bulge and disc colours has shown that in both spirals and S0s, the discs are bluer than the bulges \citep{Bothun_1990,Peletier_1996,Hudson_2010, Head_2014}, suggesting that disc galaxies have more recent star formation activity at larger radii \citep{deJong_1996_2} or higher metallicities in their nuclear regions \citep{Beckman_1996,Pompei_1997}. This method has also revealed negative colour gradients within both the bulges and discs of disc galaxies \citep{Terndrup_1994,Peletier_1996, Mollenhoff_2004, Michard_2000,Kannappan_2009,Head_2014}. Since negative colour gradients imply that redder light is more centrally concentrated within these components than bluer light, such trends imply the presence of increasingly older or more metal-rich stellar populations at smaller radii within these galaxies.

Traditionally, multi-waveband bulge--disc decomposition has been carried out on each image independently \citep[e.g.][]{Pahre_1998a,Pahre_1998b,Ko_2005,LaBarbara_2010,Lange_2015}. While this approach works in general, it is susceptible to poor fits in cases of low signal-to-noise, which can affect the analysis of the results. Some groups have tried to combat this issue. For example, \citet{Lackner_2012} used the decomposition parameters obtained from fitting the \textit{r}-band image of each galaxy to constrain all the fitting parameters except the flux in all other wavebands. However, this approach still gives no information about how the sizes and orientations of each component vary with wavelength. An alternative solution was used by \citet{Simard_2002, Simard_2011}, who were able to apply a simultaneous galaxy profile fit to 2 images, and \citet{Mendel_2014}, who showed similar results for different pairs of waveband combinations. These simultaneous fits were constrained such that only the total flux, Bulge-to-Total light ratio (B/T) and background level were allowed to vary between bands, and the images at each waveband were fitted simultaneously with the \textit{r}-band image, which is the deepest image. As a result, while they provide a better fit to the bulge and disc components, the fits are dominated by the \textit{r}-band parameters.

As a result of these issues, the `MegaMorph' project has developed \textsc{GalfitM}\footnote{http://www.nottingham.ac.uk/astronomy/megamorph/} \citep{Haeussler_2013}, a modified version of \textsc{Galfit}\footnote{https://users.obs.carnegiescience.edu/peng/work/galfit/galfit.html} \citep{Peng_2002,Peng_2010} that can decompose an arbitrary number of multi-waveband images simultaneously.  \citet{Vika_2014} found that multi-waveband decompositions with \textsc{GalfitM} improves the reliability of measurements of the component colours, and thus their colour gradients. However, estimates of the ages and metallicities of stellar populations derived from colours alone are highly degenerate \citep{Worthey_1994b}, and can also be affected by dust reddening \citep{Disney_1989}. Therefore, to better understand the star-formation histories across galaxies, the addition of spectroscopic information is vital.

A spectroscopic bulge--disc decomposition technique was developed by \citet{Johnston_2012} and \citet{Johnston_2014}, in which long-slit spectra along the major axes of S0s were decomposed into bulge and disc components. These studies fitted the one-dimensional light profiles as a function of wavelength, and then used integration of these profiles to obtain the global bulge and disc spectra for each galaxy. A different approach to decomposing long-slit spectra of S0s was used by \citet{Silchenko_2012}, who first decomposed SDSS images of each galaxy, and used the bulge and disc effective radii to determine the bulge and disc dominated regions of the galaxy. They took a spectrum for the bulge at a radius of 0.5~bulge effective radii, and subtracted off a disc spectrum that was measured from the disc-dominated region and scaled according to the bulge-to-total light ratio from the photometric decomposition. All three studies found that the bulge regions of S0s contain systematically younger and more metal rich stellar populations, but with long-slit spectra alone it is impossible to determine whether these young stellar populations are distributed throughout the bulge, or are concentrated in the centre of the disc. 

With the introduction of wide-field integral-field spectroscopic instruments, such as VLT/MUSE \citep{Bacon_2010}, and surveys, such as the Calar Alto Legacy Integral Field spectroscopy Area \citep[CALIFA, ][]{Sanchez_2012} and the SDSS-IV Mapping Nearby Galaxies at Apache Point Observatory \citep[MaNGA, ][]{Bundy_2015}, it is now possible to apply bulge--disc decomposition to three-dimensional spectroscopic data. One such study has successfully separated the bulge and disc stellar populations of S0s in CALIFA kinematically (Tabor et~al, in prep). We have developed a complementary approach called Bulge--Disc Decomposition of IFU datacubes, or \textsc{buddi},  which uses \textsc{GalfitM} to decompose the light profiles of galaxies in an integral field unit (IFU) datacube into multiple components, giving both the integrated spectra and the two-dimensional datacube for each component. 

In this paper, we present this new method and give an indication of its potential by applying the technique to a small sample of galaxies from the MaNGA survey. Section~\ref{sec:MaNGA_overview} gives and overview of the MaNGA survey and data, and Section~\ref{sec:Decomposition} describes the decomposition technique. Section~\ref{sec:Tests} looks at various tests that were carried out to ensure that the decomposition of the datacubes is reliable and consistent, and Section~\ref{sec:stellar_pops} describes the initial stellar populations analysis of a small number of galaxies. Finally, our conclusions are presented in Section~\ref{sec:Summary}

\section{Summary of MaNGA Data}\label{sec:MaNGA_overview}
The decomposition technique presented in this paper was tested on commissioning data observed as part of the MaNGA \citep{Bundy_2015} survey, observed in March and June 2014 with the Sloan 2.5m telescope \citep{Gunn_2006}. MaNGA is part of the fourth-generation Sloan Digital Sky Survey (SDSS-IV) that started on July 1st 2014 for six years, sharing dark time during that period with the eBOSS Survey \citep{Zhao_2016}. 

The MaNGA survey aims to study the internal stellar and gas kinematics and stellar population structures of a sample of $\sim$10,000 galaxies in a redshift range of 0.01~$<$~z~$<$~0.15. Up to 17 galaxies can be observed simultaneously using a series of hexagonal IFUs of different sizes plugged into a plate of 7~deg$^{2}$. The IFUs range in size from 19 fibres (12~arcsec diameter on sky) to 127 fibres (32~arcsec), and there are a further 12 mini-bundles of 7 fibres used to observe standard stars for flux calibration \citep{Yan_2016}, and 92 single fibres for sky subtraction. Each fibre has a diameter of 120 microns (2~arcsec diameter on sky), and feeds into the dual beam BOSS spectrographs \citep{Smee_2013}. 67\% of the total sample of galaxies are in the Primary sample, and are observed with the appropriately-sized IFU to ensure spatial coverage out to 1.5~effective radii in the \textit{r}-band. The Secondary sample, consisting of the remaining 33\%, will have spatial coverage out to 2.5~effective radii. Both samples cover the same mass range, but the Secondary sample is on average at a slightly higher redshift ($\langle z \rangle~=~0.045$ versus $\langle z \rangle~=~0.030$ for the Primary sample). The IFUs provide a continuous wavelength coverage between 3600 to 10300~\AA, with a spectral resolution of R$\sim$1400 at 4000~\AA\ to R$\sim$2600 at 9000~\AA\ \citep{Drory_2015}. 

With a typical integration time per field of 2-3~hours, a signal-to-noise ratio (S/N) of 4-8~\AA$^{-1}$ per fibre can be reached in the outskirts of Primary sample MaNGA galaxies, using an assumed luminosity in those regions of 23~AB~mag arcsec$^{-2}$. Each field is observed in multiples of three 15~minute exposures using a three-point dither pattern to  ensure full spatial coverage between the fibres and to obtain a uniform point spread function \citep[PSF, ][]{Law_2015}.

The data used in this paper were observed with the March and July 2014 commissioning plates, 7443 and 7815 respectively, using the 127, 91 and 61-fibre IFUs. Throughout this paper, the galaxies will be referred to by their plate number-IFU number combination. For example, galaxy 7443-12701 was observed with plate 7443 and the first of the 127~fibre IFUs. The sample covers the redshift range 0.017~$<$~z~$<$0.14, and covers a wide range of galaxy morphologies. The data was reduced using version v1.5.1 of the MaNGA Data Reduction Pipeline \citep[DRP, ][]{Law_2016}, and the reduced and flux calibrated MaNGA fibre spectra were combined to produce the final datacubes. The pixel scale in the datacubes was set to 0.5~arcsec, the PSF was found to have a full width at half maximum (FHWM) of 2.5~arcsec, and the IFUs have fields-of-view between 22~arcsec and 32~arcsec across. The datacubes use a logarithmic wavelength solution, providing a wavelength coverage of 3.5589 to 4.0151 (in units of logarithmic Angstroms), with a step size of 10$^{-4}$ dex over 4563 spectral elements \citep{Law_2016}.

\begin{figure*}
  \includegraphics[width=0.9\linewidth,bb=0 0 550 350]{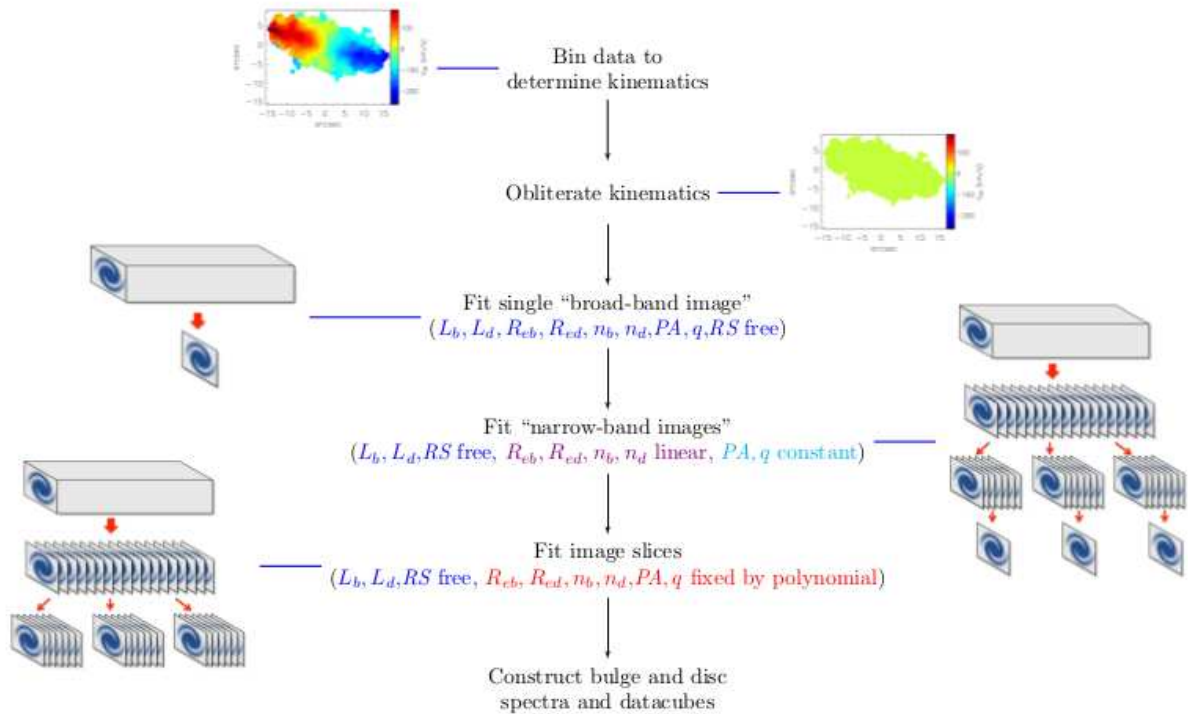}
  \caption{Overview of the steps to decompose an IFU datacube with \textsc{buddi}. The free parameters are the luminosity ($L$), effective radius ($R_{e}$), S\'ersic index ($n$), position angle (${PA}$), axes ratio ($q$) and residual sky value ($RS$), and the suffixes correspond to the bulge ($b$) and disc ($d$) components. The colours of each parameter identify how it varies with wavelength - dark blue means it's completely free with wavelength, purple is allowed to vary according to a linear polynomial, light blue can be different for different components, but remains constant with wavelength, and red is held fixed according to the best fit polynomial from the previous step.
 \label{decomp}}
\end{figure*}

\section{\textsc{buddi}: Bulge--Disc Decomposition of IFU datacubes}\label{sec:Decomposition}

One can see an IFU datacube of a galaxy as simply a stack of images at each wavelength step. If the S/N and spatial resolution is sufficiently high, features, such as the bulge, disc, spiral arms etc., can be distinguished in each image. In such cases, it is possible to decompose the datacube of the galaxy wavelength-by-wavelength to spectroscopically separate the light from each component. If however, the S/N is lower, and especially when several components with different colours are present in the galaxy, this separation is not as easy in each individual wavelength and an approach is needed that uses all information in the cube/mulitwavelength information simultaneously. Such data is ideally set up for image decomposition using \textsc{GalfitM}, a modified version of \textsc{Galfit3} \citep{Peng_2002,Peng_2010} that is able to fit multi-waveband images of galaxies simultaneously to produce one wavelength-dependent model \citep{Haeussler_2013}.  \textsc{GalfitM} fits user-defined degrees of Chebychev polynomials to the variable galaxy parameters to produce a single, consistent, wavelength-dependent model of a galaxy. These polynomial fits have the additional benefit that the fit to each image is constrained by the fits of the rest, thus boosting the signal-to-noise of the dataset over that of any individual image within the set. As a result, individual images with lower signal-to-noise only minimally affect the wavelength dependent fit \citep{Vika_2013}. 

Ideally, one would run the IFU datacube through \textsc{GalfitM} directly in order to keep the full galaxy information for the fitting process. However, due to the number of images/wavelengths involved, this would need a) a lot of memory, as the entire cube, model etc has to be kept in memory at any given time and b) a lot of CPU time, making this approach impossible in practice. If one wants to derive the full spectrum of the bulge and disc of a galaxy, each new wavelength adds another 2 free parameters for the magnitudes of each component, even if the values in all other parameters are somehow kept the same. With more than 4000 wavelength steps in the case of a MaNGA datacube, it becomes obvious why this is a futile task. Instead, one has to come up with a different, intermediate approach. In the development of \textsc{buddi}, we decided on a 4 step fitting process in \textsc{GalfitM}, in which some further pre-preparation is included. After this preparation, we first fit a broad-band image, in which all wavelengths have been combined, then a set of narrow band images, in which the full datacube has been binned in the wavelength direction in order to get a manageable number of images. Finally we return to the full wavelength resolution by fitting the individual wavelength steps in blocks of 30 in order to derive the full spectra for each galaxy component. This last step could in principle also be done individually, but we prefer this approach as it keeps a lower number of images and files on disc and is easier to handle. The outcome, however should in principle be identical (for more details of the differences between \textsc{Galfit} and \textsc{GalfitM}, see Bamford et al,~in prep). An overview of this process to decompose a galaxy datacube is shown in Fig.~\ref{decomp}, as developed in this work using MaNGA data. The individual steps are explained further in this section.

\subsection{Step 1: Data Preparation/Obliterating kinematics}\label{sec:Decomposition1}
The first step was to measure and obliterate the kinematics across the galaxy in order to obtain a homogeneous datacube, where each spaxel in each image slice measures the light at the same rest-frame wavelength and at the same position within any spectral feature at that wavelength. This step ensures that no artificial gradients in any line strengths are added due to the radial velocity of the stellar disc in highly inclined galaxies, or due to the narrowing of the spectral features further from the bulge dominated region. While decomposition is possible without obliterating the kinematics, it would lead to very broad spectral features in the final decomposed spectra, reducing the spectral resolution and blurring together neighbouring features.

The datacubes were spatially binned using the Voronoi binning technique of \citet{Cappellari_2003}, and the absorption-line kinematics measured using the Penalized Pixel-Fitting method (pPXF) of \citet{Cappellari_2004}. At this stage, the wavelengths of emission features were masked out to exclude them from the fits. The kinematics were then measured as offsets from the centre of the galaxy, which was taken as the point in the central region of the galaxy where the velocity dispersion peaked, or where the luminosity peaks in the case that the velocity dispersion shows a double peaked profile with radius. The line of sight velocity was corrected by shifting the spectrum to match that measured at the centre, and the velocity dispersion was corrected by convolving each spectrum with the appropriate Gaussian to bring it up to the maximum measured within that galaxy. 

It is important to note at this stage that if strong emission lines are present in the spectrum that do not trace the same kinematics as the absorption features, they will not be corrected to the same level of accuracy as the absorption spectrum. Large misalignments in the kinematics, particularly in cases of counter-rotating gaseous discs, would lead to contamination of the decomposed spectra by artificially broadened emission features. Such an effect can be seen at $\sim$5577~\AA\ in Fig.~\ref{fig:decomposed_spectra} where the spectrum contains significant residuals of a bright sky line. While modelling the emission lines is beyond the scope of this work, it is expected that in future work strong emission features will be modelled and subtracted reliably at this stage of the fitting process to reduce their effects on the final decomposed spectra.

\subsection{Step 2: Fitting one broad-band image}
Having obliterated the kinematics within the datacube, the galaxy can be fit with \textsc{GalfitM}. The first step was to median stack the entire datacube into a single, broad-band or white-light image of the galaxy, and to decompose it with \textsc{GalfitM}.  In this first step, however, where only one broad-band image is used for the fit, these multi-band features are technically not needed and the normal version of \textsc{Galfit} could be used instead. For consistency and in order to only rely on one code, we have chosen to use \textsc{GalfitM} even in this first step.

After creating the broad-band image, a mask was created to identify the zero-value pixels in the image that are outside of the hexagonal MaNGA IFU field of view. This mask was used in this step and the subsequent steps to ensure that only those pixels with information were included in the fits.

The initial fit was carried out using a single S\'ersic profile and a residual-sky component, and a PSF image was included for convolution. The PSF profile was created by median stacking the broad-band \textit{griz}-band PSF images for that datacube, which are provided as part of the MaNGA data products. Although no large influence on the final fitting result was recorded in previous tests, the starting parameters in \textsc{GalfitM} are somewhat important to the fit, especially in multi-component decompositions. As we were unable to find a general automatic solution for deriving those starting values,  we estimated starting parameters by eye for this first step.

After fitting the broad-band image with \textsc{GalfitM}, the residual image created by subtracting the model from the original image was used to visually identify whether the starting parameters are reasonable and if an additional component would be necessary to improve the fit. In cases where the latter was true, the fit was repeated with an additional S\'ersic profile. For ease of analysis in this study, the more extended component has been assumed to be the disc, while the more centrally concentrated component has been labelled as the bulge. As a further check of the initial parameters for this fit, we have carried out a similar fit with the same starting parameters on photometric data of the same galaxy (see section~\ref{sec:SDSS_images}). 

Since the model light profiles created by \textsc{GalfitM} are smooth, any non-smooth features within the galaxy, such as spiral arms or bright knots of star formation within the disc, will not be included in the model. In the presence of very bright or very large structures, it can happen that they are included in the fit to the bulge or disc, thus artificially enhancing the luminosity of the component. In such cases, they must therefore be masked from the fit.

\subsection{Step 3: Fitting narrow-band images}\label{sec:decomposition_3}
After fitting the broad-band image, the datacube was binned in the wavelength direction to produce a series of narrow-band images. The purpose of this step is to derive the polynomials of most galaxy parameters that describe how they change with wavelength. These polynomials can then be used in the last fitting step to derive the full spectra of the galaxy components. As long as the number of images used here is sufficiently high, it is not necessary to carry out the fitting of these polynomials by using all images at all wavelengths, saving huge amounts of CPU time in the process. In the case of the MaNGA data presented here, it was found that using 30 narrow-band images provided a good balance between S/N, the number of data points for the polynomial fits, and the time needed for \textsc{GalfitM} to fit that many images. 

The Chebyshev polynomials used by \textsc{GalfitM} are arbitrarily normalized over a range of -1 to 1, which is mapped to the spectral range of the input data. In order to ensure that the best-fit polynomials for the narrow-band images cover the full spectral range of the IFU cube, the first and last narrow-band images must be the image slices at the start and end wavelengths of the spectral range covered, instead of e.g. the center of the first and last narrow band images. Otherwise the polynomial coefficients produced by this second step can not be used for the full spectral range in the last step. Therefore, the 30 narrow-band images consist of 28 median-stacked images plus the first and last image slices of the desired wavelength range.

For simplicity, the binning of the median-stacked images was applied by separating the remaining datacube over the desired wavelength range into 28 equal regions, and stacking the individual images within each region to create the narrow-band images. Since the datacubes were binned logarithmically in the wavelength direction, each of these stacked images were created from equal numbers of image slices, or channels. In the case of the MaNGA datacubes, using a wavelength range of 3700~$<~\lambda~<$~10300~\AA, each narrow-band image was created by median stacking $\sim$160 image slices. Therefore, any strong emission features present in the spectra had only a small effect on the final narrow-band image. It was not necessary to exclude the regions around bright emission features when creating the binned images since \textsc{GalfitM} uses polynomials to define the fit parameters, and thus can fit through these affected images using the information from the rest of the wavelength range.

With multi-waveband fits, it is important to account for the wavelength variation in the PSF profile, and so for each narrow-band image a PSF image at that wavelength was included. The PSF images corresponding to each narrow-band image were created by interpolating between the four broad-band \textit{griz} PSF images.

The images were fit with \textsc{GalfitM}, using the best-fit model from the broad-band image (from step 1) as the initial parameters for the multi-waveband fit. For the fits of these narrow-band images, only the luminosities of the bulge and disc ($L_{b}$ and $L_{d}$ respectively) and the residual-sky background were left completely free. 
Due to the wealth of the information present in the multi-band images and the constraints between images due to the polynomials used, we found that this freedom was not a problem as long as the field-of-view contained even a small region of sky.

\begin{figure*}
  \includegraphics[width=0.9\linewidth]{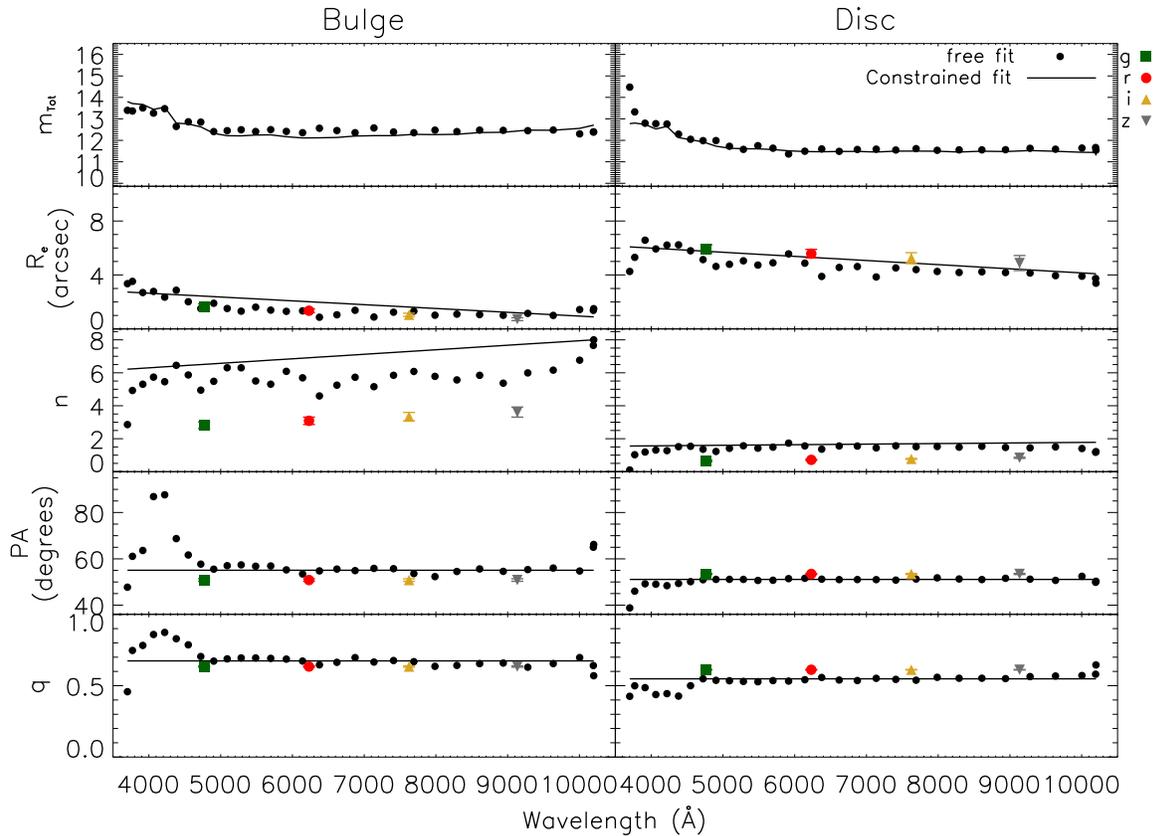}
  \caption{The decomposition parameters obtained for the narrow-band images of galaxy 7443-9102, showing the results for both the free (dots) and the constrained fits (solid line) using \textsc(GalfitM). The two columns represent the results from the bulge (left) and disc (right), and from top to bottom the parameters are the integrated magnitude, the effective radius, the S\'ersic index, the position angle and the minor/major axis ratio. Overplotted as coloured squares, circles and triangles are the decomposition parameters obtained after fitting the SDSS \textit{griz}-band images with \textsc{GalfitM}.
\label{fig:free_fits}}
\end{figure*}
%decomp_paramams.pro

The effective radii of the bulge and disc ($R_{eb}$ and $R_{ed}$ respectively) and their S\'ersic indices ($n_{b}$ and $n_{d}$) were constrained to vary linearly over the wavelength range. The position angle (\textit{PA}) and axis ratio ($q$) of each component were assumed to be constant with wavelength while being allowed to be different between the two components. These restrictions on the polynomials for the variables were selected to improve the fit over the full wavelength range, compensating for individual image slices with higher levels of noise or contamination by sky lines, while also allowing enough freedom to look for colour gradients within the bulge and disc, as measured from the gradients in their effective radii \citep[see e.g. ][]{Johnston_2012}. 

\begin{figure*}
\centering
\begin{minipage}{.45\textwidth}
  \centering
  \includegraphics[width=1\linewidth]{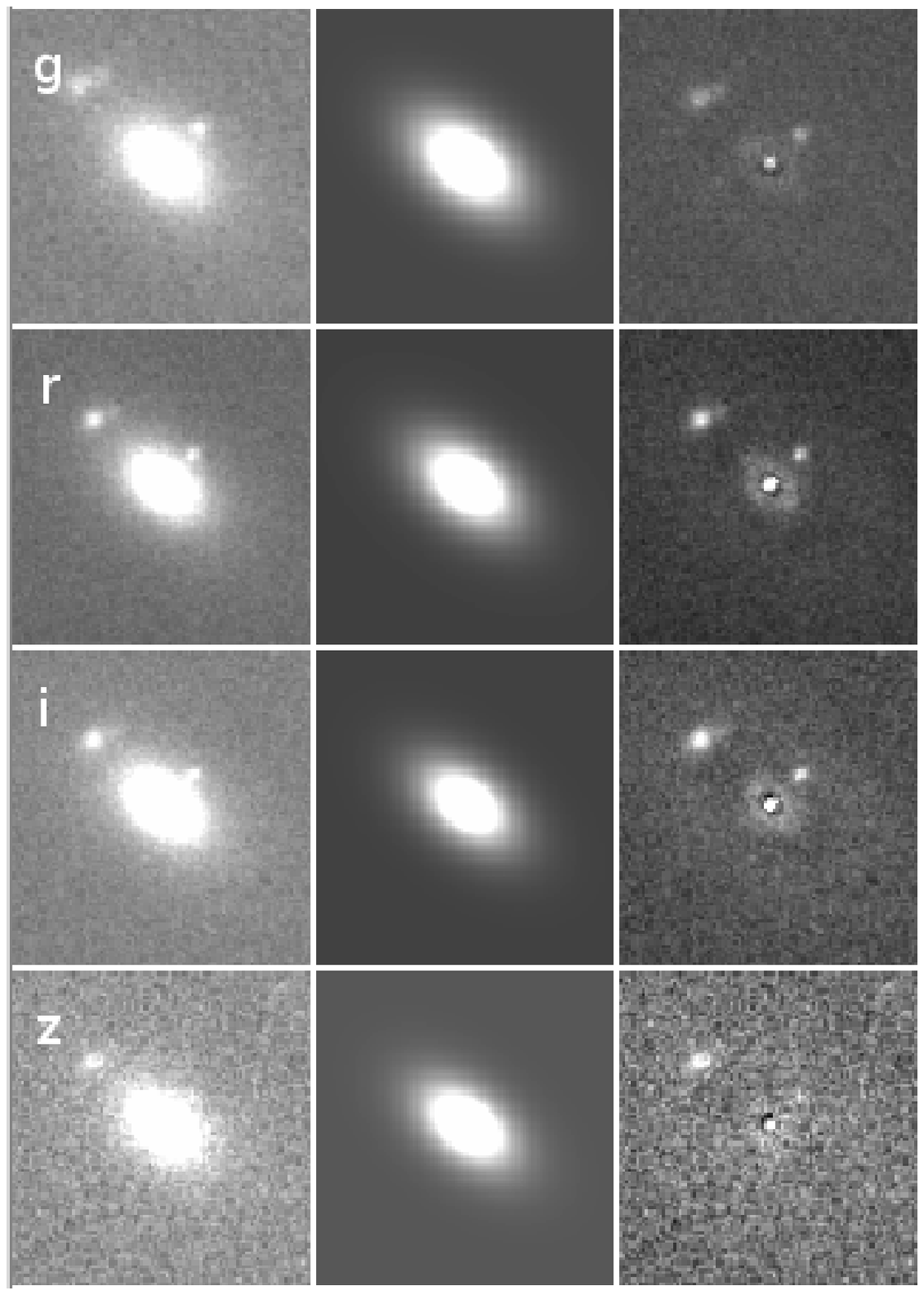}
 \caption{Comparison of the decompositon of the SDSS griz images. The columns from left to right show the original griz images for galaxy 7443-9102, the best fit model composed of two S\'ersic profiles, and the residual images after subtracting the model from the original image. The colour distribution of all images are linear, with a range of 99.5\% of the total range in pixel values. The scaling of the residual images has been adjusted to best show any features present.}
 \label{decomp_SDSS}
\end{minipage}%
\hspace{.05\linewidth}
\begin{minipage}{.45\textwidth}
  \centering
  \includegraphics[width=1\linewidth]{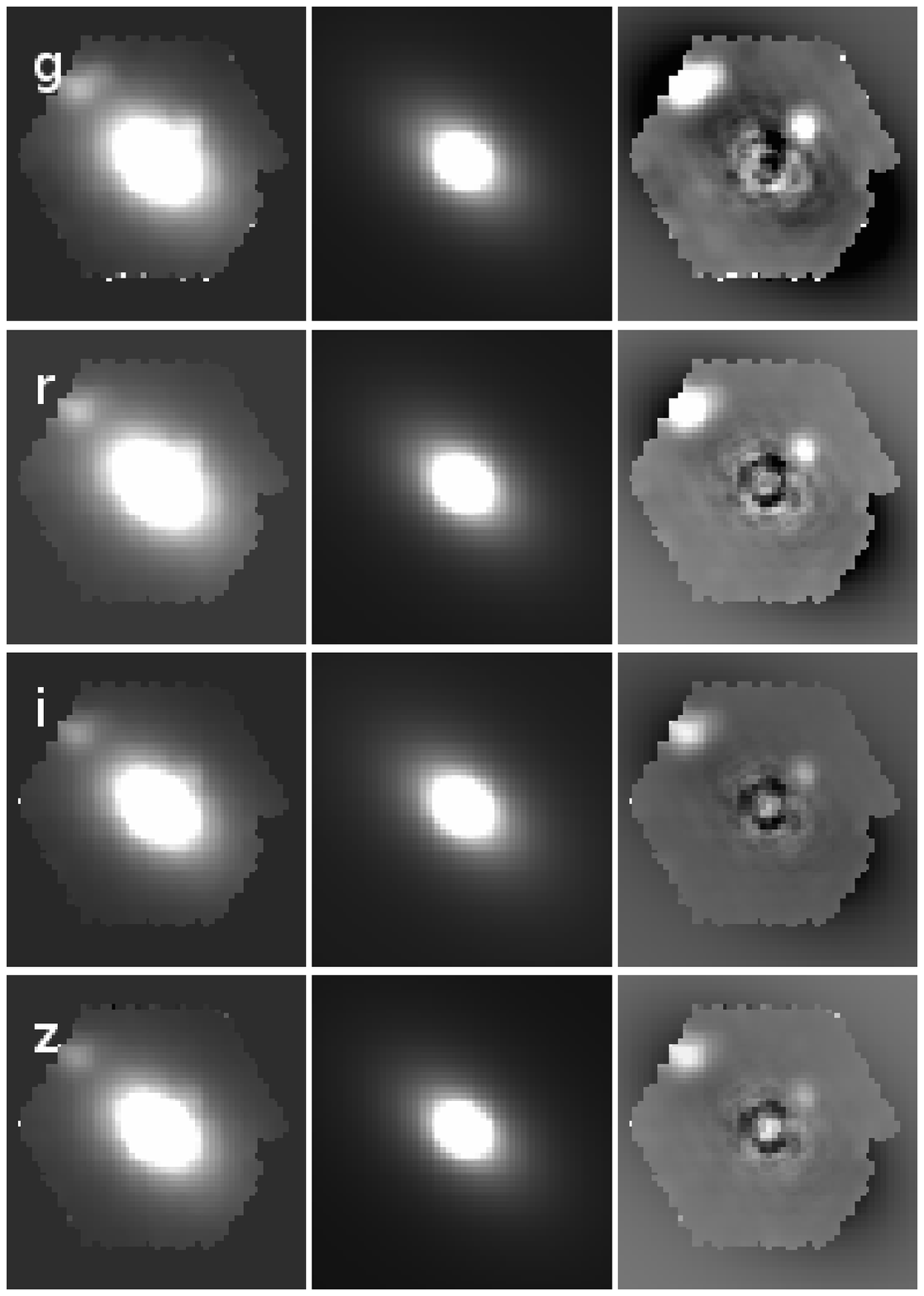}
  \caption{As for Fig.~\ref{decomp_SDSS} but for the MaNGA broad-band images. It can be seen here that the stacking of the images from the circular fibres smoothes and broadens any features, an effect which is particularly noticeable in the object on the bottom left of the image. The colour distribution of all images are linear, with a range of 99.5\% of the total range in pixel values. The scaling of the residual images has been adjusted to best show any features present.}
 \label{decomp_MANGA}
\end{minipage}
\end{figure*}

A comparison of the constrained and free fits, in which all parameters were given full freedom, to the narrow-band images of galaxy 7443-9102 are given in Fig.~\ref{fig:free_fits}. As a further example, the results from fitting SDSS \textit{griz} images in the same way are also overplotted- see Section~\ref{sec:SDSS_images} for more details. It can be seen that the constrained parameters fit well to the unconstrained parameters, thus providing a reliable smooth variation with wavelength. Not all galaxies are best fit with the constraints listed here \citep[see e.g.][]{Kelvin_2012}, and in such cases it may be necessary to allow more freedom in the models by increasing the orders of the polynomials. In this galaxy, the largest differences between the constrained and free fits are in the bulge S\'ersic indices. The most likely reason for this discrepancy is likely to be the difference in the spatial resolution- the imaging data has a spatial resolution of $\sim$1.3~arcsec FWHM while the MaNGA images have a resolution of 2.5~arcsec- leading to blurring together of different features and artificial broadening of compact sources. Such broadening effects can be seen in the foreground objects in Figures~\ref{decomp_SDSS} and \ref{decomp_MANGA}, which show examples of the best fits to the SDSS and MaNGA \textit{griz}-band images for this galaxy. The residual images, which were created by subtracting the best-fit model from the SDSS images, show a bright, compact source of light at the centre of the galaxy that has not been included in the fit. In the MaNGA data however, this feature appears much fainter and broader in the residual images, typically accounting for about 1.9$\%$ of the total light in the same region in the original images compared to $\sim$4.2$\%$ in the SDSS \textit{griz}-band images. This result suggests that the light from this central bright component has been blurred with the bulge light in the MaNGA data, leading to an increase in the bulge S\'ersic index due to the brighter centre. 

It can also be seen that, in some cases, the free parameter fits for the first and last images are very different to the rest of the results for that parameter. This effect is simply due to these images being the individual image slices at the start and end wavelengths of the spectral range, and thus they have lower S/N than the rest of the dataset and result in poorer fits. It has also been found that the S/N at extreme wavelength ranges in MaNGA data tend to be lower due to lower throughput in these regions \citep[see e.g. Fig~4 of][]{Yan_2016}. In the constrained fits, these images in particular are supported by neighbouring images, leading to a much more reliable fit even in these images. 

%originally 7815-9102

\subsection{Step 4: Fitting individual wavelength images}
Having decomposed the narrow-band images and obtained the polynomials describing the wavelength dependence of the component parameters, it is now possible to decompose the datacube wavelength-by-wavelength. Due to the computing power needed to decompose the whole datacube in one go, it is again necessary to split it up into chunks of, say, 30 image slices. Technically, this step could also be carried out wavelength-by-wavelength using \textit{galfit} single band fitting. However, we chose to fit the blocks of images with \textit{GalfitM} to reduce the number of intermediate files created (i.e. one file created per 30 images instead of per image) and to maintain consistency buy using the same fitting software throughout the whole process. With the polynomials for all structural parameters already known from the previous step, it is unnecessary to allow any freedom in any of those parameters. Doing so would only introduce artefacts at the wavelengths where the blocks of images meet. For this final decomposition, only the bulge and disc luminosities and the residual sky value were allowed complete freedom, while all the remaining parameters were held fixed at the values determined at each wavelength through calculation from the polynomial derived by the narrow-band fit. While such restrictions do limit the information that can be obtained, such as the measurement of line strength gradients within each component, they constrain the fits sufficiently that each image slice can be decomposed in a consistent way.

\begin{figure*}
  \includegraphics[width=0.95\linewidth]{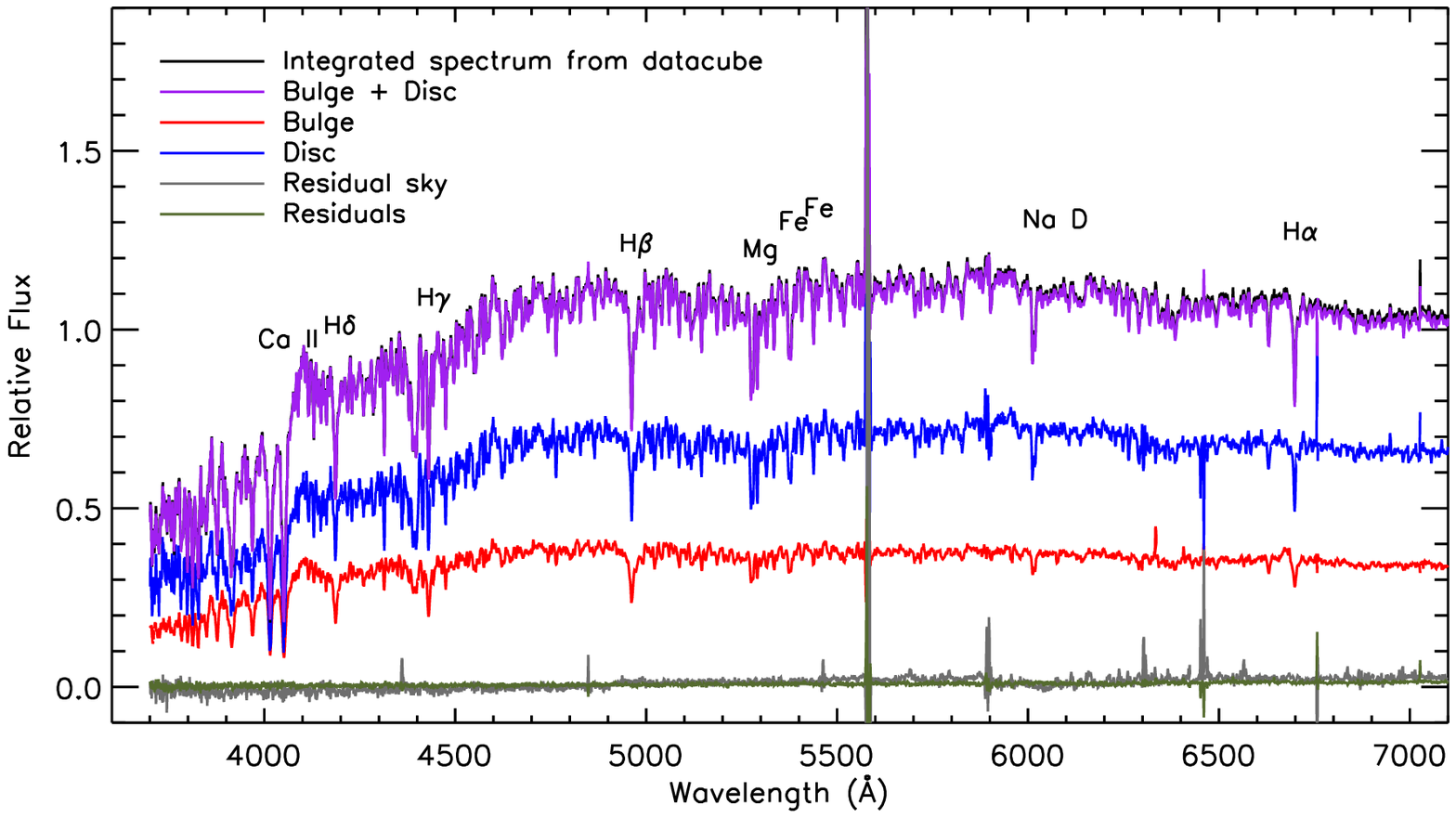}
  \includegraphics[width=0.95\linewidth]{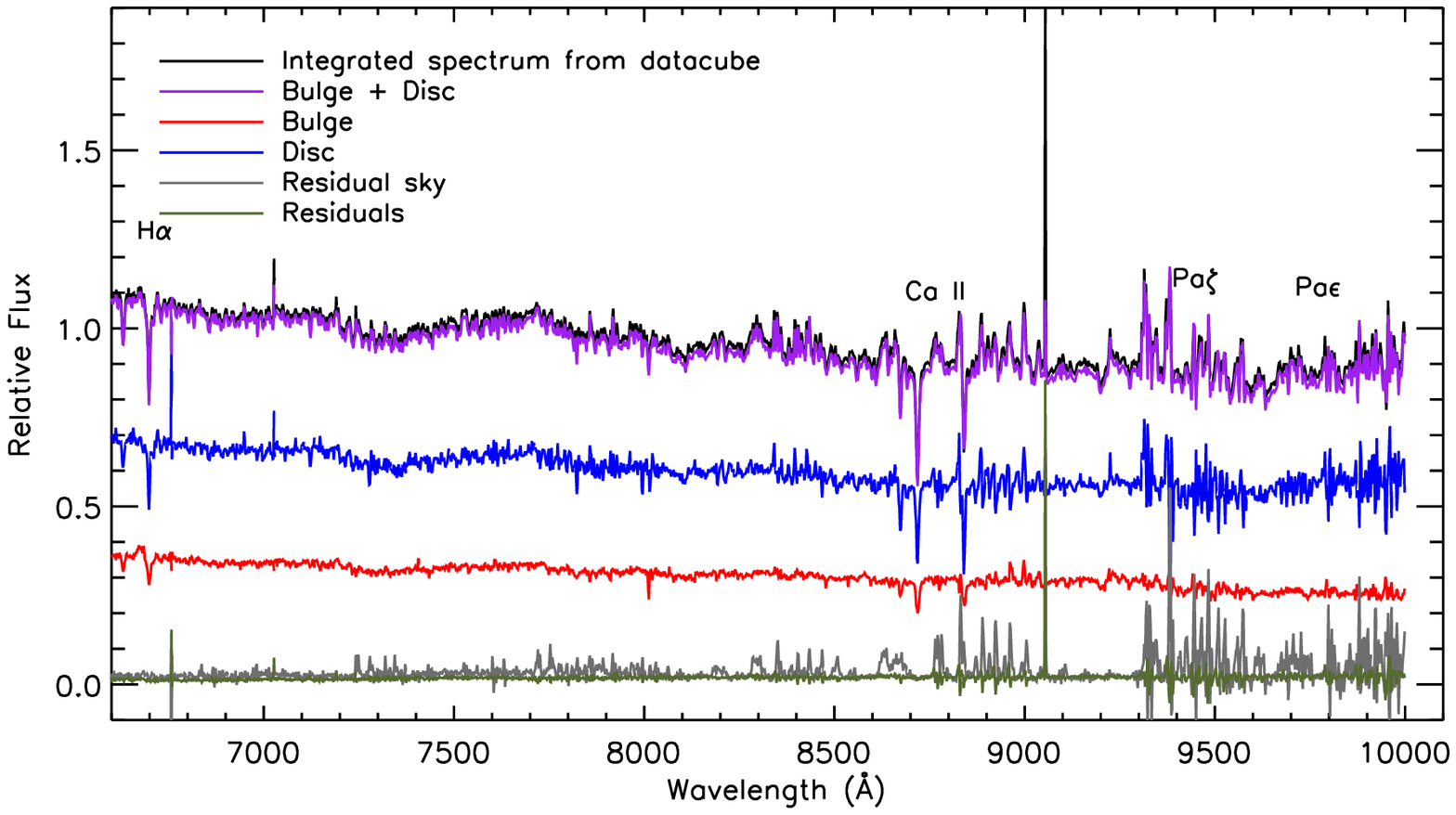}
  \caption{Decomposed one-dimensional bulge and disc spectra for galaxy 7443-12701 in red and blue respectively. The black line shows the original integrated spectrum, the over plotted purple line is the combined bulge+disc+residual sky spectra, and the grey line gives the spectrum of the residual sky component. Finally, the green line shows the residuals after subtracting the combined bulge+disc+residual sky spectra from the original. The top plot shows the optical spectra, and the bottom plot shows the near-IR spectra with a 500~\AA\ overlap in the wavelength coverage. The feature at 5577~\AA\ in the top plot is the [OI] sky line.
\label{fig:decomposed_spectra}}
\end{figure*}
%IFU_decomposition->result_visualiser_2

Finally, from the resulting luminosities at each wavelength, the decomposed spectra can be reconstructed in two ways. The first is to simply plot the total luminosity of each component as a function of wavelength to produce a one-dimensional integrated spectrum for that component, from which the mean, global stellar populations and star-formation histories can be measured. An example of the decomposed one-dimensional bulge and disc spectra for galaxy 7443-12701 is shown in Fig.~\ref{fig:decomposed_spectra}, along with the original spectrum, the best fit from the combined bulge+disc+residual sky spectrum overplotted on the original, the residual sky component included in the fit, and finally the residuals after subtracting the best fit from the original. The majority of the regions of poor fits and higher levels of noise in the fits correspond to regions of non-zero residual-sky background and poorly subtracted strong sky lines. While the data reduction pipeline does a good job of sky subtraction in the MaNGA datacubes using the dedicated sky IFUs, generally to better than Poisson error estimates, the sky subtraction is not perfect, especially in regions of strong sky lines \citep{Law_2016}. For example, part of the strong [OI] sky line at 5577.4~\AA\ has been included in the bulge and disc models, with the remainder of the light at that wavelength appearing in the residuals. Similarly, the increased residuals at redder wavelengths coincides with the increase in the residual-sky background level and the number of sky lines in this region. The skylines should in principle mostly contribute to the residual-sky background measured, not the actual profile luminosity. It is reassuring that we do find the residual-sky background elevated at and around skylines, as it shows that the simultaneous background estimation in these fits works, despite the lack of any clean `sky' pixels that are usually needed to reliably determine the residual-sky background level.

 The second technique is to use \textsc{GalfitM} to create images of each component at each wavelength from the decomposition parameters. These images can then be stitched together to produce IFU datacubes representing purely the light from each component. These datacubes can be used to visualise how each component varies with wavelength, and potentially to study stellar population gradients within each component. Such gradients were detected by \citet{Johnston_2012} in the decomposition of long-slit spectra through peaks or troughs in the plot of the effective radii with wavelength at the wavelength of the spectral feature. However, these features could only be detected in the decomposition of IFU datacubes if the effective radii of each component were unconstrained with wavelength during the fit. However, the S/N of the MaNGA data is lower than the spectra used by \citet{Johnston_2012}, in which the S/N was $\sim$53~per \AA\ at the centre of the galaxy, and so the unconstrained fits produced very noisy results. One possible solution would be to perform the decomposition over the spectral region of interest only (e.g. around a spectral line) and to choose an appropriate polynomial with which to constrain the effective radii.
  
 In addition to creating the smooth decomposed bulge and disc datacubes, it's possible to also create a residuals datacube by subtracting the best fit model from each image slice. Since the residual images subtract the smooth model from the non-smooth galaxy, they can be used to isolate any non-smooth features such as spiral arms or bright star-forming knots. By stacking the images into a datacube, the spectra of these features can be extracted, thus allowing a separate analysis of their star-formation histories.

\subsection{Feasibility in Terms of Computing Power}\label{sec:speed}
The four-step process outlined in this section allows the decomposition of a MaNGA datacube to be carried out on a desktop or laptop computer in a reasonable length of time. For example, using a Mac laptop with 8GM RAM, the entire decomposition process from original datacube to decomposed spectra can be carried out over about $\sim$9~hours for a galaxy using a double S\'ersic profile over the full wavelength range of MaNGA. The slowest step of the process is the final fitting of the individual wavelength images, for which the time required depends on the complexity of the model used, the number of images/wavelength range to be fitted etc. In the example above, the individual image fits took $\sim$7.5~hours to complete. Since \textsc{GalfitM} only uses one core at a time for the fits, it is possible to speed up the fitting process for large samples of galaxies by running multiple fits simultaneously on powerful computers with multiple parallel cores. Similarly, the fit could be sped up by using the narrow-band images to derive the polynomials over the full wavelength range available, but then only applying the fits to individual images over a limited wavelength range to cover only the region of interest for the analysis.

\section{Reliability Tests}\label{sec:Tests}
IFU data generally have smaller fields-of-view and poorer spatial resolution than imaging data typically used for galaxy fitting. As a result, the small field-of-view limits the view of the outskirts of the galaxy, where the bulge light may dominate the light profile again, and the poor spatial resolution blurs the signal from the different components in the inner regions. In the case of MaNGA, the field-of-view is limited to show only out to $\sim$1.5 or 2.5 effective radii, and for all but the most edge-on galaxies there are few spaxels with no galaxy light with which to quantify poisson variations in the residual sky background, especially around bright skylines. Therefore, before application to the MaNGA datacubes, the decomposition technique outlined in Section~\ref{sec:Decomposition} was tested to identify the reliability of the decomposed spectra and the limitations of such analysis. This section outlines a series of tests carried out to determine the feasibility of fitting galaxies observed with IFU instruments, specifically looking at the effect on the best fit due to the small field-of-view and the poorer sampling compared to imaging data.

\subsection{Decomposition of simulated galaxies}\label{sec:BD_params}

The largest MaNGA IFU has a field of view of 32.5 arcseconds across, and the IFUs are assigned to each galaxy to capture their light out to either 1.5 or 2.5~R$_{e}$. Since the decomposition of photometric data with software such as  \textsc{Galfit} and \textsc{GalfitM} generally uses a larger field of view, the first set of tests were designed to identify the reliability of the decomposition for the various fields of view of the MaNGA galaxies. One reliable way of testing such parameters is to create and decompose a series of simulated galaxies in which the structural properties and stellar populations are known \citep[see e.g.][]{Simard_2002,Haeussler_2007}.

The first step in producing these simulated datacubes was to create an image of the galaxy with a de Vaucouleurs bulge and exponential disc profile. The images were created with \textsc{Galfit} using a range of values for the luminosity, effective radius, position angle and axes ratio for each component. In each model galaxy, the bulge component was modelled as the de~Vaucouleurs profile with a smaller effective radius than the disc, with a total magnitude fainter than the disc by up to one magnitude and with the same position angle. Additionally, each image was convolved with a PSF image with a FWHM of 2.5~arcsec. A series of 96 simulated galaxy datacubes were created, consisting of four model galaxies, each observed at three inclinations (nearly edge-on, intermediate and nearly face-on) with two fields of view (1.5 and 2.5~R$_{e}$) as seen in the four largest MaNGA IFUs (127, 91, 61 and 37-fibres).

To create the datacube of the simulated galaxies from these images, \textsc{Galfit} was used to create images of the bulge and disc for each model. The amount of light in each spaxel from the bulge and disc was taken to be the fraction of light from each component in that spaxel. The datacube was then created by coadding the bulge and disc spectra in the correct proportions for each spaxel. 

The bulge and disc spectra were taken from the MILES stellar library \citep{Sanchez_2006}, which contains stellar spectra of known age and metallicity covering a wavelength range of 3525-7500~\AA\  at 2.5~\AA\ (FWHM) spectral resolution \citep{Falcon_2011, Beifiori_2011}. The ages and metallicities of the stellar spectra used to create the simulated galaxies spanned the ranges of 2~$<$~Age~$<$~15~Gyrs and -1.31~$<$~[M/H]~$<$~0.22 respectively. We assigned younger bulges and older discs to two model galaxies, and older bulges and younger discs to the other two galaxies, where one galaxy for each scenario was assumed to have a single stellar population within the bulge and disc while the other galaxy contained multiple stellar populations within each component. Additionally, three of the bulges contained higher metallicities than their corresponding discs, while the fourth bulge was more metal poor. For simplicity, it was assumed that there were no age or metallicity gradients within either component. Due to the combination of different stellar populations selected for each component within each galaxy and the corresponding variety of shapes of their spectra, the Bulge-to-Total (B/T) light ratio varied both between each simulated galaxy and with wavelength in these galaxies.

Finally, a noise component was added to the simulated datacubes. The noise was assumed to be Poissonian and independent of wavelength, and was applied to each image slice within the datacubes.

\begin{figure*}
  \includegraphics[width=1\linewidth]{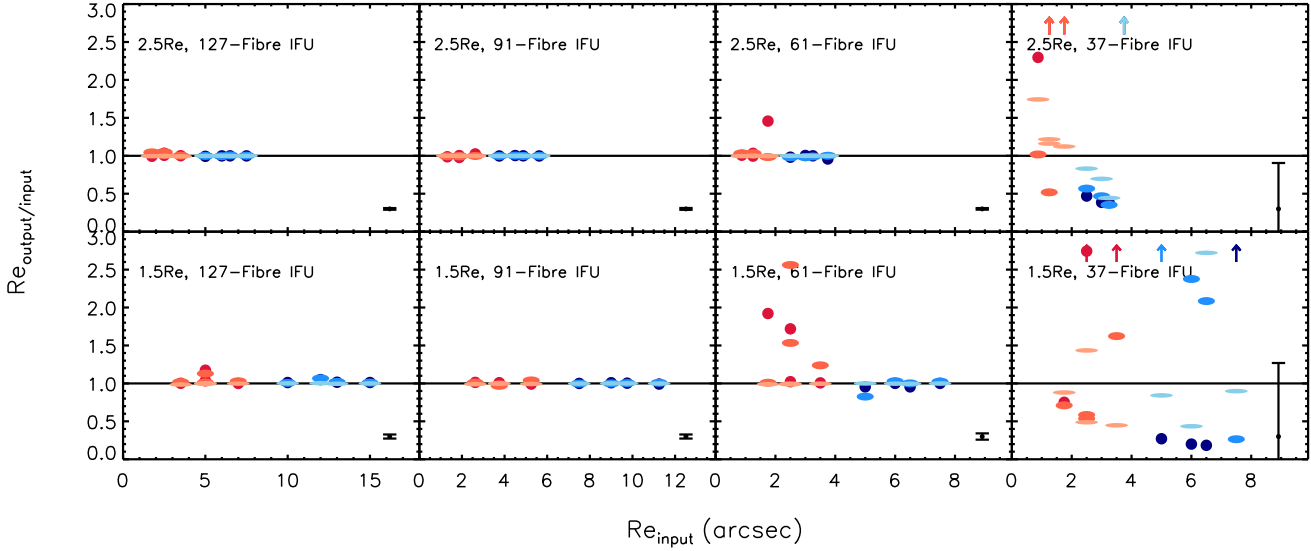}
  \caption{A comparison of the input and output effective radii obtained by \textsc{GalfitM} for a series of simulated datacubes. The top row represents galaxies in which 2.5~Re were included in the field of view while the bottom row includes galaxies in which only 1.5~Re are covered, and the columns show the results for the 127, 91, 61 and 37-fibre IFUs from left to right. The red and blue points represent the bulge and disc components respectively, while the ellipticity describes the orientation to the line-of-sight- circles are almost face-on galaxies while the fainter ovals represent galaxies with higher inclinations.
    \label{Re_comparison}}
\end{figure*}
%input_output_comparisons_3/4.pro
\begin{figure*}
  \includegraphics[width=1\linewidth]{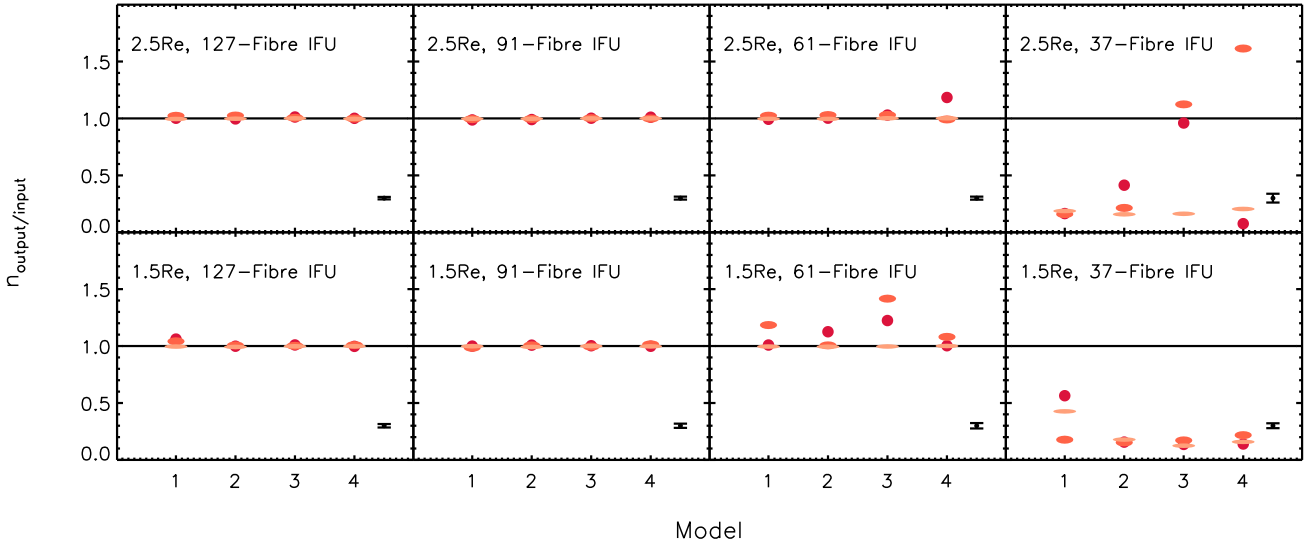}
  \caption{As for Fig.~\ref{Re_comparison}, but showing the results for the S\'ersic indices of the bulge component. The x-axis in this case is simply the model number since each simulated galaxy consisted of an exponential disc and a de~Vaucouleurs bulge.
    \label{n_comparison}}
\end{figure*}

The simulated datacubes were decomposed as described in Section~\ref{sec:Decomposition} using an exponential plus S\'ersic model. The results for the bulge and disc effective radii and the bulge S\'ersic indices are given in Fig.~\ref{Re_comparison} and Fig.~\ref{n_comparison} respectively, plotted against the input parameters for each IFU and field of view. The exception is the bulge S\'ersic index, which is plotted against model number since the input parameters are the same in each case. The mean uncertainty for each setup is plotted in the bottom right, and represents the statistical uncertainties calculated in the fit by \textsc{GalfitM}. It is important to remember that the uncertainties calculated by \textsc{GalfitM} are based on a galaxy that can be fitted well with the number of components included in the model and that only has Poissonian noise. Such scenarios are very rare, and so these uncertainties should be considered a lower limit \citep{Haeussler_2007}. It can be seen that in the larger IFUs, the fit parameters obtained from the decomposition are relatively close to the input values, while the smaller IFUs show significantly more scatter. The worst fits were obtained with the 37-fibre IFU, which has a field-of-view of 17~arcsec in diameter. In this case, even with the simplified models used here, \textsc{GalfitM} struggled to find a good fit with a two-component model, and the resultant spectra appeared too noisy to use.

%line_index_comparison_3.pro
\begin{figure*}
  \includegraphics[width=1\linewidth]{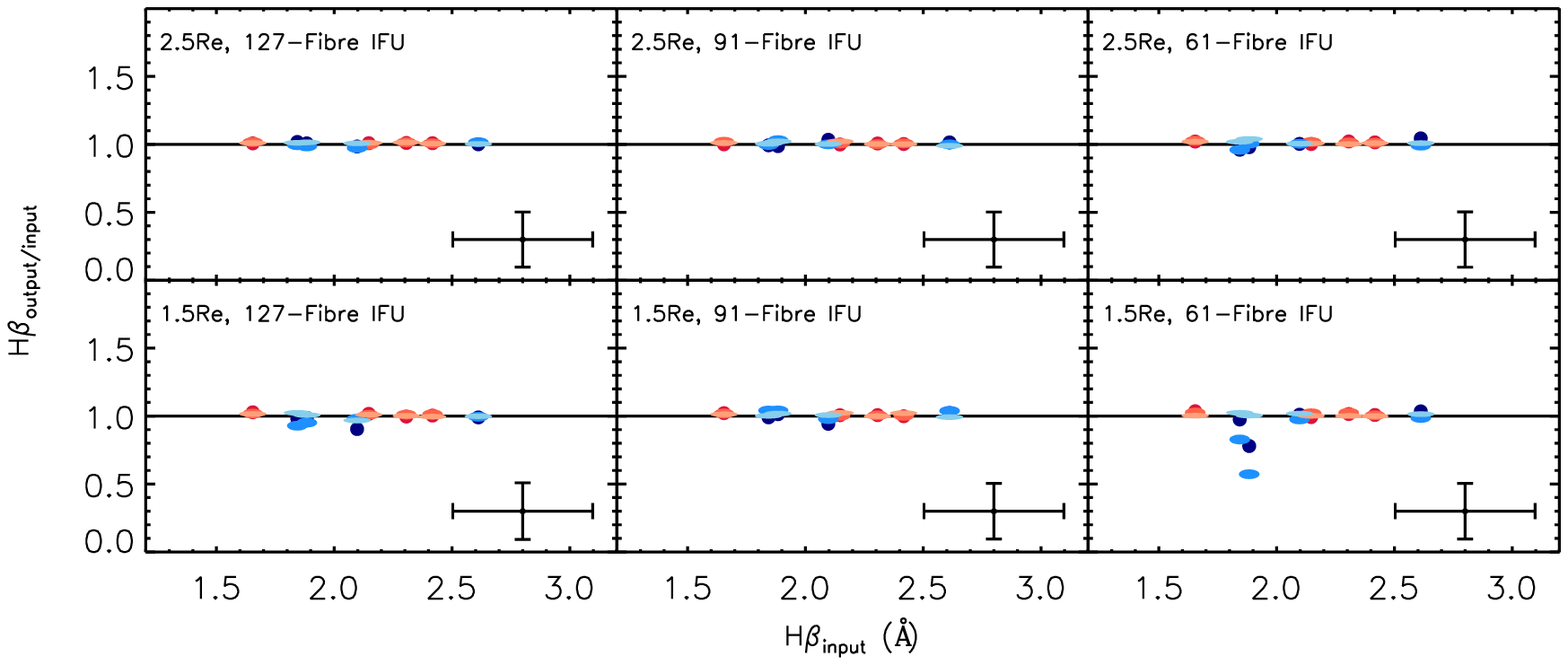}
  \caption{As for Fig.~\ref{Re_comparison}, but showing the results for the strength of the H$\beta$ absorption features.
\label{Hb_comparison}}
\end{figure*}

\begin{figure*}
  \includegraphics[width=1\linewidth]{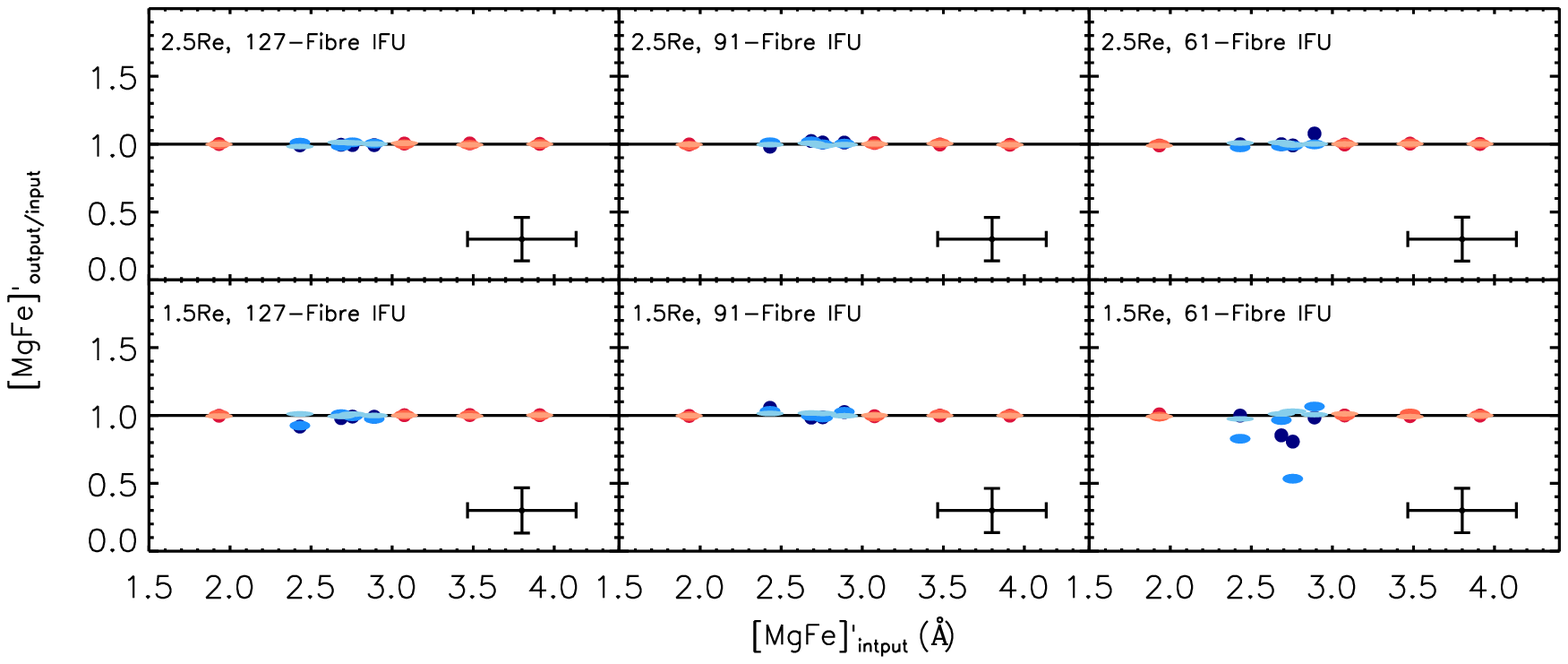}
  \caption{As for Fig.~\ref{Re_comparison}, but showing the results for the strength of the combined [MgFe] absorption features.
    \label{MgFe_comparison}}
\end{figure*}

The increasing scatter in the decomposition parameters towards smaller IFUs is most likely due to the smaller field of view of the IFUs leading to a smaller number of pixels being used to fit the galaxy and the residual-sky. Additionally, the 2.5~arcsec FWHM spatial resolution leads to blurring of the bulge and disc light, with the result that \textsc{GalfitM} is unable to fully disentangle the high and low S\'ersic index components in the bulge-plus-disc model used in this example. These trends however allow us to restrict our sample selection within MaNGA to those IFUs that provide a sufficiently large field-of-view, and suggests caution when using the decomposed spectra in regions outside of the IFU field-of-view where the fits are poorest.

The next step was to determine how much of an effect the scatter in the decomposition parameters have on the resultant bulge and disc spectra and the stellar populations analysis. The results for the 127, 91 and 61-fibre IFU datacubes can be seen in Fig.~\ref{Hb_comparison} and Fig.~\ref{MgFe_comparison} for the H$\beta$ and [MgFe]$'$ line strengths respectively. The results for the 37-fibre IFU datacubes are omitted here since the fits were so poor that the decomposed spectra appeared too noisy to be useful. The line strengths were measured using the Lick/IDS index definitions, in which the strength of the spectral feature is measured relative to a pseudo-continuum calculated from the level of the spectrum in bands on either side \citep{Worthey_1994, Worthey_1997}.  The combined metallicity index, [MgFe]$'$, is defined as
\begin{equation} 
	\text{[MgFe]$'$}=\sqrt{\text{Mg}b\ (0.72 \times \text{Fe}5270 + 0.28 \times \text{Fe}5335)},
	\label{eq_MgFe}
\end{equation}
\citep{Gonzalez_1993,Thomas_2003}. The scatter in the line strength measurements is seen to increase for a combination of smaller IFUs and galaxies observed only out to 1.5~R$_{e}$. Additionally, the scatter in the line strengths appear larger for the discs than for the bulges. This effect is most likely due to the loss of light from the outskirts of each galaxy due to the small field of view of the MaNGA IFUs. The mean uncertainty in the line index measurements can be seen in the bottom right of each plot, as calculated through a series of numerical simulations with synthetic error spectra for each galaxy. 
Since the scatter is generally within the error limits, it can be deduced that uncertainties in the decomposition parameters do not affect the strength of the features in the decomposed spectra significantly.

\subsection{Comparison with decomposition of SDSS images}\label{sec:SDSS_images}
The above test assumes that the galaxies being decomposed can be modelled almost perfectly with a simple two-component model. In reality however, galaxies have more complicated morphologies due to the presence of spiral arms, bars, thin and thick discs etc, which can affect the fit when the galaxy is modelled as a two-component system. Such additional components may cause further issues when decomposing IFU data due to the combination of the relatively small field of view, poorer spatial resolution and low signal-to-noise per image compared to photometric data. A further complication of the MaNGA data is that the circular fibres and the dither pattern used can cause features within the galaxy to become smoothed and slightly broadened once the data is stacked. For example, while the expected observational seeing is $\sim1.5$~arcsec, the FWHM of the reconstructed PSF after fibre sampling and stacking the dithered images will be $\sim2.5$~arcsec \citep{Bundy_2015}. However, such smoothing may also be useful for the fitting since it removes any sharp, small-scale features that may distort the fit.

To better understand the extent of the limitations when fitting IFU data, both imaging and IFU broad-band images of the galaxies observed with the March 2014 and July 2015 commissioning plates (plates 7443 and 7815 respectively) and the 127, 91 and 61-fibre IFUs were decomposed and the results compared. The imaging data was obtained from the SDSS DR7 \citep{Abazajian_2009} \textit{griz}-band images, and mosaicked together with the \mbox{\textsc{Montage}} software\footnote{http://montage.ipac.caltech.edu/} to produce a large field of view ($\sim$8~arcmin centred on the galaxy) for the two-dimensional image decomposition. The MaNGA broad-band \textit{griz} images and corresponding PSF files are provided as part of the MaNGA data products included with each datacube. These broad-band images were created by using the \textit{griz}-band transmission curves to stack the relevant images in the datacube in the right ratios, and thus are a useful comparison to the imaging data.

The imaging and IFU images for each galaxy were fitted in the same way using the same starting parameters, and all fits started with a single exponential profile to represent the disc plus a residual-sky background component. In all cases, the residual-sky background was assumed to be constant over the field-of-view in each image. After each fit, the residual images, which were generated by \textsc{GalfitM} by subtracting the best-fit model from the original image, were checked by eye to identify whether the addition of another component in the inner regions would improve the fit. In cases where the residual images showed obvious over/undersubtraction of the galaxy light in the inner regions, or where dark rings/patches could be seen, the fits were repeated with an additional S\'ersic or a PSF profile to identify which combination produced the best fit. The residual images were again checked by eye and compared to the earlier fit to determine whether an improvement had been achieved. While comparing the residual images by eye is not guaranteed to determine whether the best fit has been achieved, for a small sample of galaxies it is a useful indicator as to whether one or two components provide the best fit. Similarly, the residual images can help identify whether \textsc{GalfitM} has settled on a local minimum, resulting in a poor overall fit. By checking the fits by eye in this way, it was found that good fits for the MaNGA broad-band \textit{griz} images from the 61-fibre IFU could only be achieved when using a single component. As described in Section~\ref{sec:BD_params}, the small number of spaxels in these images contributes to the poorer fits when additional components were added. Additionally, due to the IFU selection for  MaNGA galaxies to cover out to at least 1.5~effective radii, the galaxies observed with the 61-fibre and smaller IFUs are smaller galaxies in general, leading to increased blurring of the light from each component at smaller radii.

For each component, the effective radius and S\'ersic index (where applicable) were constrained to a linear polynomial over the four wavebands, the axes ratio and position angle were constrained to remain constant with waveband, and the magnitudes were allowed complete freedom. Figure~\ref{fig:free_fits} presents an example of the results for the decomposition parameters of 7443-9102 from the imaging data over plotted on the results from the decomposition of the narrow-band images described in section~\ref{sec:Decomposition}. Despite the difference in the S/N of the broad-band and narrow-band images, the fit parameters are very similar over the wavelength range. The main difference lies in the S\'ersic index of the bulge, which is likely to be due to the varied levels of blurring by the different spatial resolutions of the photometric and spectroscopic data, as described in Section~\ref{sec:decomposition_3}.

\begin{figure*}
\centering
\begin{minipage}{.5\textwidth}
  \centering
  \includegraphics[width=0.85\linewidth]{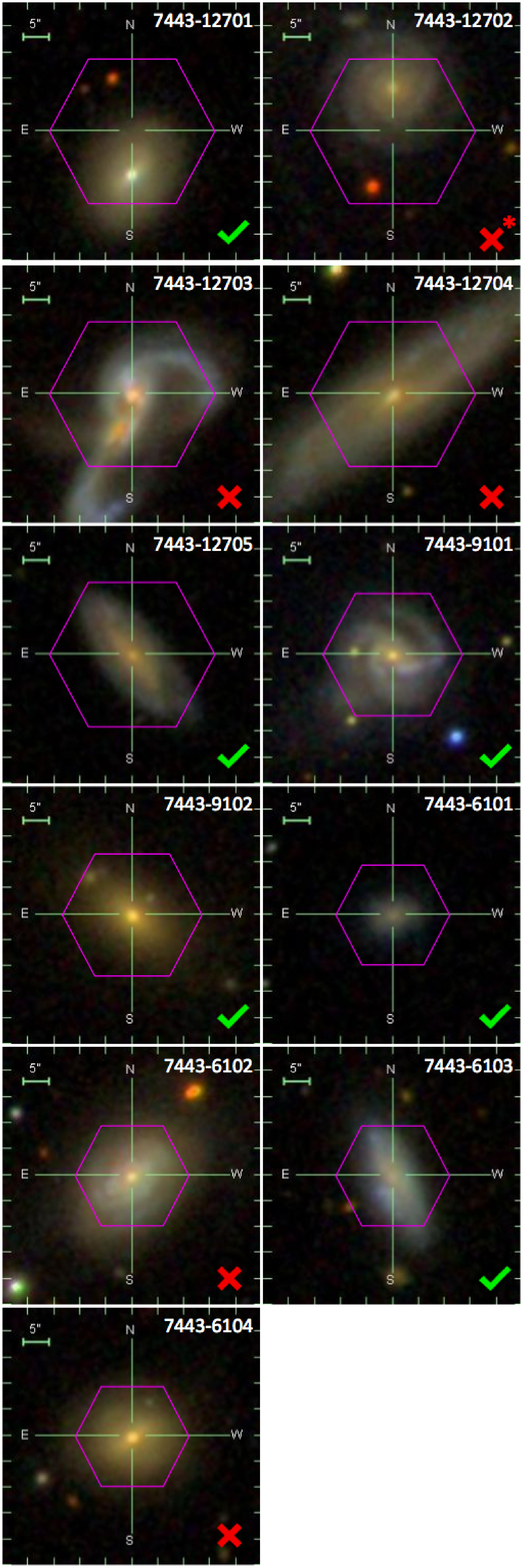}
\end{minipage}%
\begin{minipage}{.5\textwidth}
  \centering
  \includegraphics[width=0.85\linewidth]{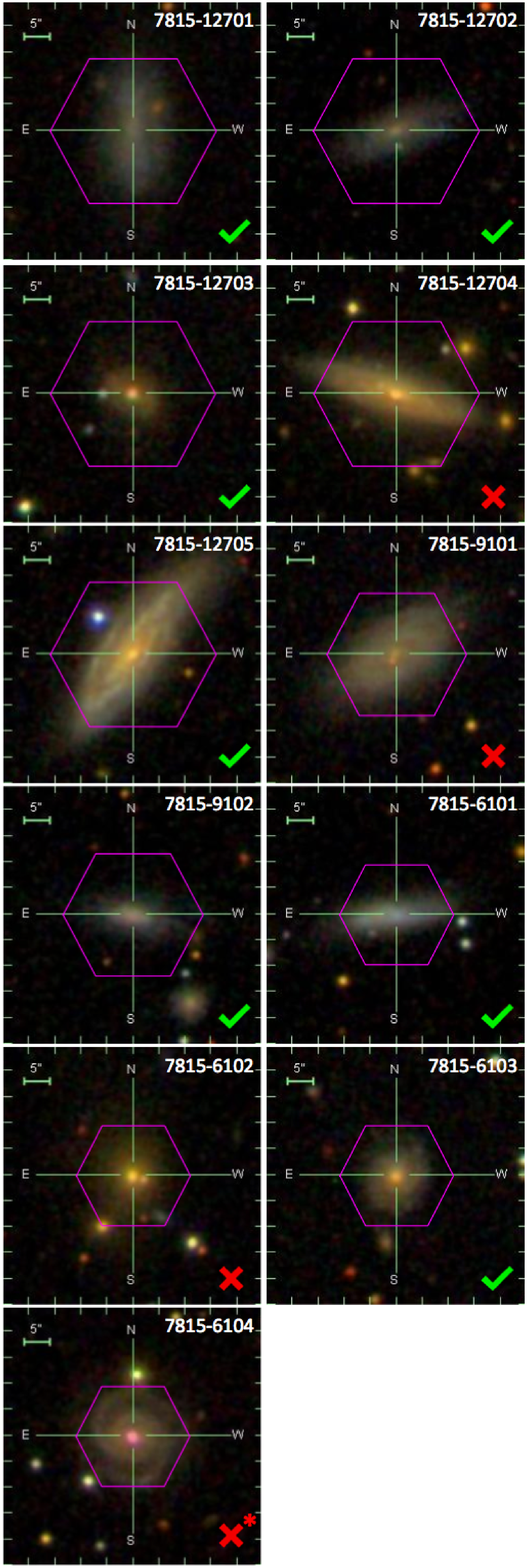}
\end{minipage}
\caption{SDSS images of the galaxies observed with the 127, 91 and 61-fibre IFUs on commissioning plates 7443 (left) and 7815 (right). The plate number-IFU number is listed at the top right of each galaxy, and the field-of-view of the IFU is over plotted. The ticks and crosses in the bottom right of each image show whether  or not that galaxy could be fitted with either an exponential or an exponential-plus-S\'ersic profile, and the crosses with asterisks show the two galaxies in which fits could only be obtained when a S\'ersic profile was used to model the disc instead of an exponential.}
\label{fig:images}
\end{figure*}

Figure~\ref{fig:images} presents each galaxy that was analysed, along with the IFU field-of-view and whether it was successfully fitted with the exponential or exponential-plus-S\'ersic profile. Of the 22 galaxies analysed in this way, 13 galaxies were successfully fitted in both data sets, indicating a success rate of $\sim$~59~$\%$. Reasons for the unsuccessful fits include galaxies that were not simple disc-only or bulge-plus-disc systems, including one case where the IFU was centred on two interacting galaxies, or where the galaxies were too face-on and the residual-sky background couldn't be fit reliably, especially where the spatial coverage extended out to only 1.5~R$_{e}$ of the galaxy.

\begin{figure*}
\centering
\begin{minipage}{.5\textwidth}
  \centering
  \includegraphics[width=0.9\linewidth]{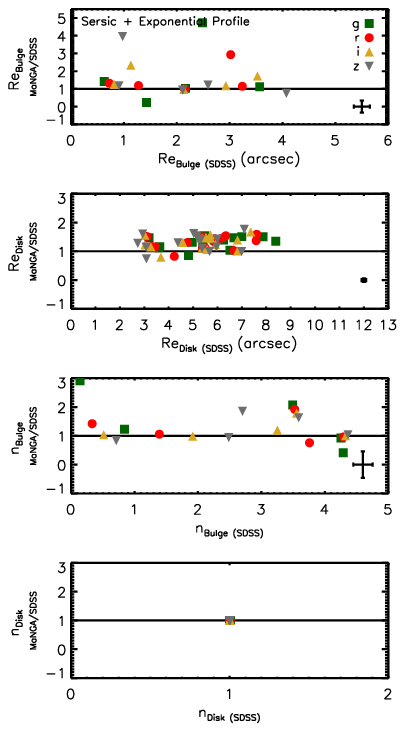}
\end{minipage}%
\begin{minipage}{.5\textwidth}
  \centering
  \includegraphics[width=0.9\linewidth]{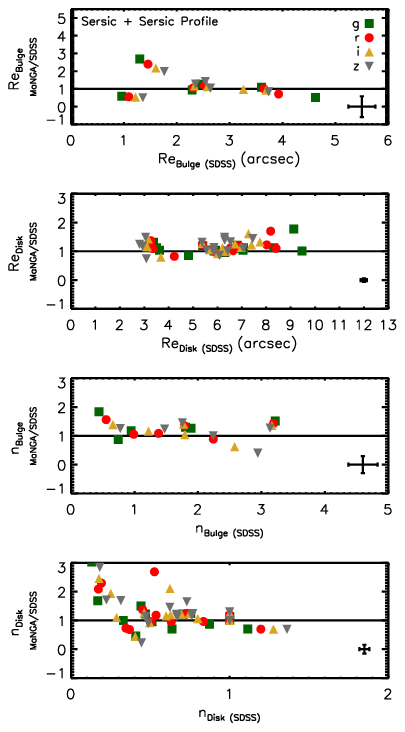}
\end{minipage}
\caption{Comparison of decomposition parameters for SDSS and MaNGA broad band images where the decompositions assume an exponential (left) and S\'ersic (right) disc. From top to bottom are the effective radii for the bulge and disc, then the S\'ersic indices for the bulge and disc. The \textit{griz}-band images are represented as green squares, red circles, yellow upward-pointing triangles and grey downward-pointing triangles respectively. In all cases, galaxies were fitted with an exponential disc with either an additional S\'ersic or PSF profile added where this  additional component was found to improve the fit. The colours refer to the band that each measurement corresponds to.}
\label{SDSS_MANGA2}
\end{figure*}

A comparison of the bulge and disc fit parameters obtained for the imaging and MaNGA broad-band images is given in Fig.~\ref{SDSS_MANGA2}, where the different colours and symbols correspond to the different wavebands for each galaxy that could be successfully modelled. The mean statistical uncertainty for each parameter over all four bands as calculated by \textsc{GalfitM} is plotted in the bottom right of each plot. It can be seen that the effective radius of the disc appears in general overestimated slightly in the MaNGA data relative to the imaging data, particularly in the case of a fixed exponential disc.  The effective radii of the bulges of two galaxies in the MaNGA data were also measured to be  larger than in the imaging data by a factor of more than 3. At this point it important to remember that the hexagonal MaNGA field of view is smaller than the square image size, and so the regions outside of the MaNGA field of view are masked out when a fit is applied with \textsc{GalfitM}. However, as can be seen in the model and residual images of Fig.~\ref{decomp_MANGA}, the best-fit model created by \textsc{GalfitM} expands into the masked region. Thus, \textsc{GalfitM} creates a best-fit model for the whole galaxy despite only having information on the inner regions. To check the effect on the fits with the outer regions masked out, the fits carried out in this section were repeated using the SDSS images with a mask restricting the field-of-view to that seen by the MaNGA data. The decomposition parameters were then compared to those obtained with the full field, and the same trends as shown in Fig.~\ref{SDSS_MANGA2} were obtained. Therefore, it is believed that the over-estimation in the effective radii of the two components in Fig.~\ref{SDSS_MANGA2} is likely to be due to this loss of light from the outer, disc-dominated regions of the galaxy and the subsequent less-accurate fits to these regions by \textsc{GalfitM}.

%image_spectra_comparisons_2_n1_4.pro
%image_spectra_comparisons_2_4.pro

Since not all galaxies have perfectly exponential discs \citep{vanderKruit_1979,Pohlen_2002, Maltby_2012}, these fits were repeated starting with a single S\'ersic profile instead of the exponential profile. With this model, 15 galaxies could be decomposed successfully, increasing the success rate to $\sim$~68~$\%$. The results given in Fig.~\ref{SDSS_MANGA2} appear to show a better correlation between the two data sets. The S\'ersic indices for the discs now show the largest scatter, again most likely due to the loss of light in the outermost regions affecting the fits, but given the small values of the disc S\'ersic indices, the variation is not so large. The disc effective radii show a smaller overestimation than when the discs were constrained to exponential profiles, and the scatter in the bulge sizes and S\'ersic indices are also lower in amplitude. Combined, these trends indicate that in some galaxies, better fits can be obtained for the MaNGA data when their discs are modelled as S\'ersic profiles as opposed to exponential profiles.

Since the IFU and imaging data have similar spatial resolutions, 2.5 and 1.3~arcsec FWHM of the PSF respectively,  the scatter in the plots in Fig.~\ref{SDSS_MANGA2} should not be significantly affected by the different resolutions. Instead, it is most likely attributed to the smoothing created by stacking the circular apertures of the fibres, and the loss of light in the outskirts of each galaxy leading to the poor fitting of the bulge and disc light in these regions. Examples of a double-S\'ersic decomposition for 7815-9102 for the SDSS and MaNGA broad-band images are shown in Figs.~\ref{decomp_SDSS} and \ref{decomp_MANGA} respectively, giving the \textit{griz} broad-band images along with the corresponding best fit models and the residual images. These images show that the features in the MaNGA data are smoother and broader with less pixel-to-pixel variations than the SDSS images due to the correlated signal and noise among spaxels compared to the more independent SDSS image pixels. This smoothing is primarily due to the PSF covering 5 pixels in the MaNGA images, while in the SDSS images the PSF covers only three pixels. Additionally, the residual images for the MaNGA data shows how the best fit to the galaxy extends beyond the MaNGA field-of-view, leading to the poor separation of bulge and disc light at large radii. However, despite these differences between the data sets, the decomposition parameters measured from the imaging and MaNGA data for this galaxy were found to be consistent.

\subsection{\textsc{GalfitM} vs. \textsc{Galfit}}\label{sec:Galfitm_Galfit}
One of the key strengths of \textsc{GalfitM} over \textsc{Galfit} is that by fitting multiple images simultaneously, the S/N of the entire data set increases over that of any individual image. Additionally, by fitting the decomposition parameters as polynomials as opposed to discrete values, a reasonable estimate of the parameters of individual images with low S/N can still be obtained.

\begin{figure*}
  \includegraphics[width=0.9\linewidth]{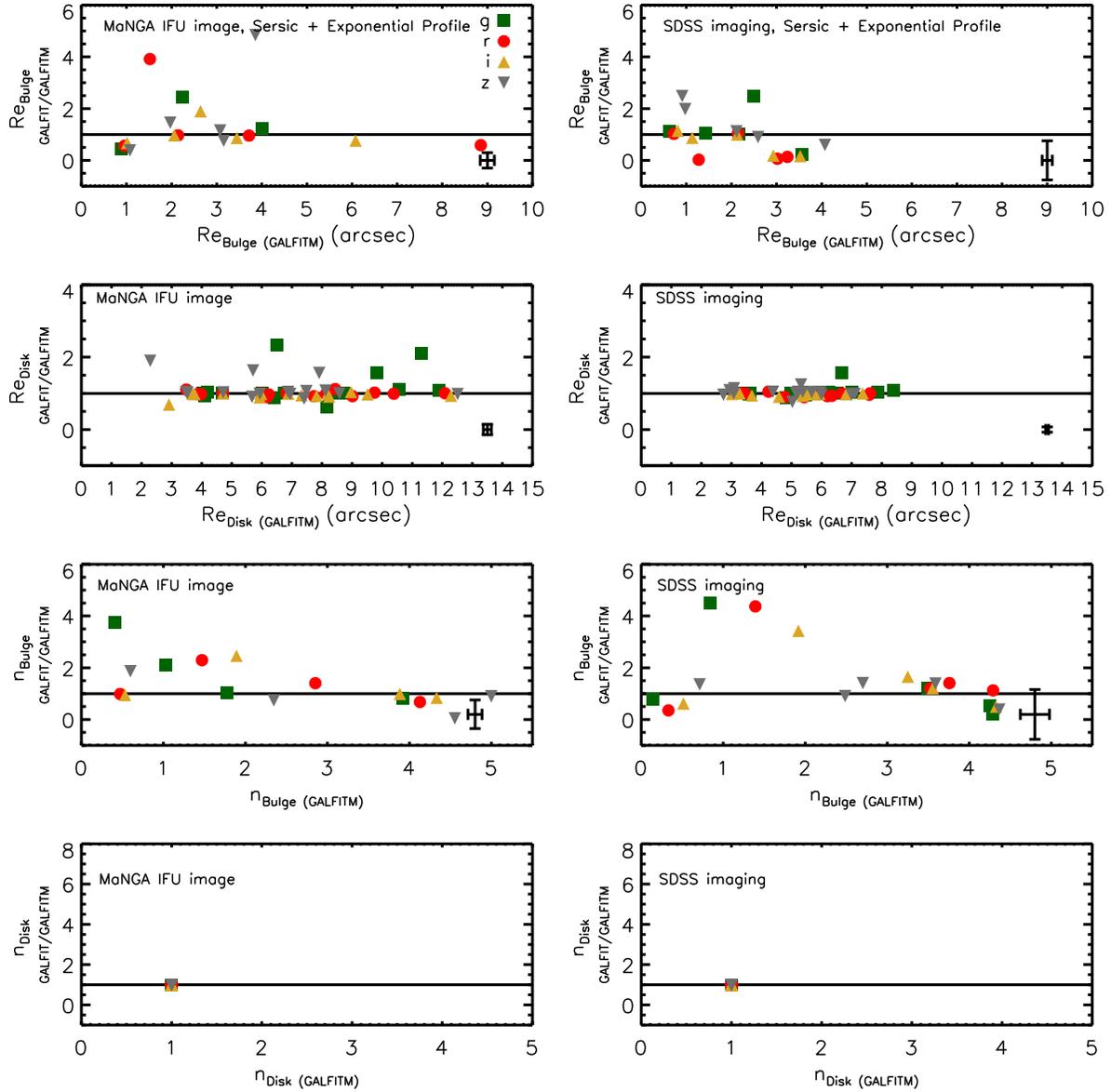}
  \caption{Comparison of decomposition parameters for SDSS and MaNGA \textit{griz} broad-band images (right and left respectively) obtained using \textsc{GalfitM} for constrained fits and \textsc{Galfit} for completely free fits. The plots from top to bottom represent the bulge and disc effective radii and the bulge and disc S\'ersic indices. In all cases, galaxies were fitted with an exponential disc with either an additional S\'ersic or PSF profile added where this additional component was found to improve the fit.
    \label{Galfit_Galfitm2}}
\end{figure*}

\begin{figure*}
  \includegraphics[width=0.9\linewidth]{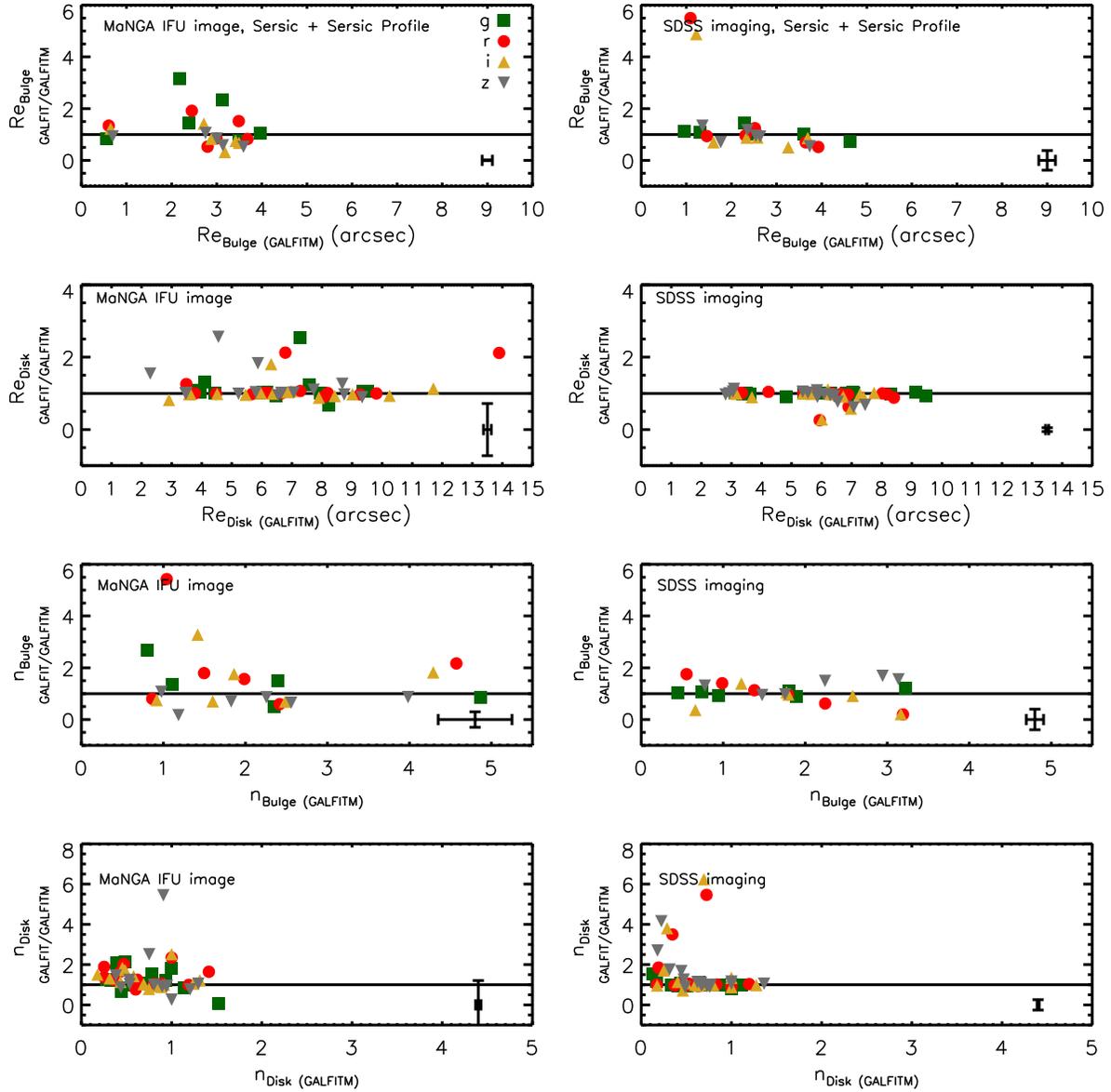}
  \caption{As for Fig.~\ref{Galfit_Galfitm2} but for the fits that assumed a S\'ersic profile for the disc.
 \label{Galfit_Galfitm1}}
\end{figure*}

To better understand how much of an improvement is achieved in the fits using \textsc{GalfitM}, the decompositions described in Section~\ref{sec:SDSS_images} were repeated using \textsc{GalfitM} with all parameters allowed complete freedom to emulate the results \textsc{Galfit} would give. The results are given in Figs.~\ref{Galfit_Galfitm2} and \ref{Galfit_Galfitm1} for the decompositions starting with exponential and S\'ersic profiles respectively for both the MaNGA and SDSS data sets.

While carrying out the free fits, it was immediately clear that without any constraints from the other multi-waveband images of the same galaxy, \textsc{GalfitM} was finding very unphysical parameters for some images. For example, some galaxies were fitted with S\'ersic indices of between 10 and 20, axes ratios of less than 0.1 or effective radii of less than half a pixel. Such poor fits consequently make up the majority of the outliers in Figs.~\ref{Galfit_Galfitm2} and \ref{Galfit_Galfitm1} and emphasise the strength of using \textsc{GalfitM} to decompose images from datacubes reliably.

The amplitude of the scatter in the results for R$_{e}$ and \textit{n} are reduced when the disc is constrained to an exponential profile, though some outliers due to poor fits are still present. However, as was discussed in Section~\ref{sec:SDSS_images}, not all of the galaxy discs are best represented by an exponential profile and by forcing this one constraint, the size of the disc and the model for the bulge may also be affected.

\subsection{Summary}\label{sec:tests_summary}
Together, the results presented in this section show that the large number of images in an IFU datacube that can be decomposed simultaneously leads to a better extraction of information, even with small fields of view and poor spatial resolution. Therefore, if the decomposition of an IFU datacube is carried out carefully and initiated with reasonable staring parameters, reliable results for the independent stellar populations within each component can be obtained.

\section{Example analysis of the decomposed spectra}\label{sec:stellar_pops}
Traditional stellar population analysis of galaxies using spectra typically involves using radially binned spectra, where each bin contains a superposition of light from each component. The spectroscopic decomposition technique outlined above produces decomposed spectra representing purely the light from each component, thus reducing the contamination from other light sources and allowing us to study their individual star-formation histories.

In this section we present example stellar population analyses for the separated bulge and disc spectra from galaxies 7443-12701 and 7443-9102 in order to demonstrate the potential of the method described in this paper. These two galaxies were selected since they are both early-type disc galaxies with no strong emission features or other significant structures such as spiral arms, and could be decomposed with a bulge-plus-disc model. Images of the galaxies, along with the field-of-view of the IFUs, are shown in Fig.~\ref{fig:images}. Both galaxies were observed as part of the Primary MaNGA sample, meaning that they were observed out to $\sim$1.5~effective radii as measured from single S\'ersic fits. 

7443-12701 has been classified as having a probability of 0.57 and 0.75 of being an early-type galaxy by \citet{Huertas-Company_2011} and \citet{Lintott_2011} respectively, and the single-fibre SDSS DR-12 data for this galaxy shows features indicative of a post-starburst spectrum. It lies at a redshift of $z~\sim$~0.020 \citep{Aihara_2011}, and was observed with a 127-fibre IFU with a field-of-view of 32.5~arcsec across. Figure~\ref{fig:images} shows that the galaxy was offset from the centre of the IFU, the purpose of which was to test the astrometric routines using the nearby star during commissioning. This offset is unusual within the MaNGA pointings, but provided a useful first test for our decomposition since the model could be fitted to the outskirts of the disc and the sky on one side. A faint bar can also be seen in this galaxy, and the effect on the fit is discussed in Section~\ref{sec:Decomposition}. 

The second galaxy, 7443-9102, has been classified as an early-type galaxy with probabilities of 0.90 \citep{Huertas-Company_2011} and 0.79 \citep{Lintott_2011}. It lies at a redshift of $z~\sim$~0.092 \citep{Aihara_2011}, and was observed with the 91-fibre IFU with a field-of-view of 27.5~arcsec. This galaxy is a typical example of an observation of a Primary sample galaxy within MaNGA since it is centered within the IFU.

Both galaxies were decomposed as described in Section~\ref{sec:Decomposition}, and the following stellar populations analysis were carried out on the decomposed one-dimensional bulge and disc spectra.

\subsection{Light-weighted stellar populations}\label{sec:LW_stellar_pops}
The light-weighted stellar populations of the decomposed spectra were measured using the hydrogen, magnesium and iron absorption line strengths as indicators of age and metallicity respectively. The line strengths were measured using the Lick/IDS index definitions, and the combined metallicity index, [MgFe]$'$ (see equation~\ref{eq_MgFe}), was used 
due to its negligible dependence on the $\alpha$-element abundance \citep{Gonzalez_1993,Thomas_2003}. The age was measured using the H$\beta$ absorption feature, corrected for contamination from emission by using the [O\textsc{iii}]$_{\lambda5007}$ emission line strength in the relation 
\begin{equation} 
\Delta(\text{H}\beta)=0.6 \times
  \text{EW}[\text{O}\text{\textsc{iii}}]_{\lambda5007}
\end{equation}
\citep{Trager_2000}, where the equivalent width of the [O\textsc{iii}]$_{\lambda5007}$ feature was measured from the residual spectrum obtained by subtracting the best combinations of stellar templates produced by \textsc{ppxf} from the original decomposed spectrum.
The uncertainties in the line index measurements were estimated from the propagation of random errors and the effect of uncertainties in the line-of-sight velocities.

\begin{figure*}
  \includegraphics[width=1\linewidth]{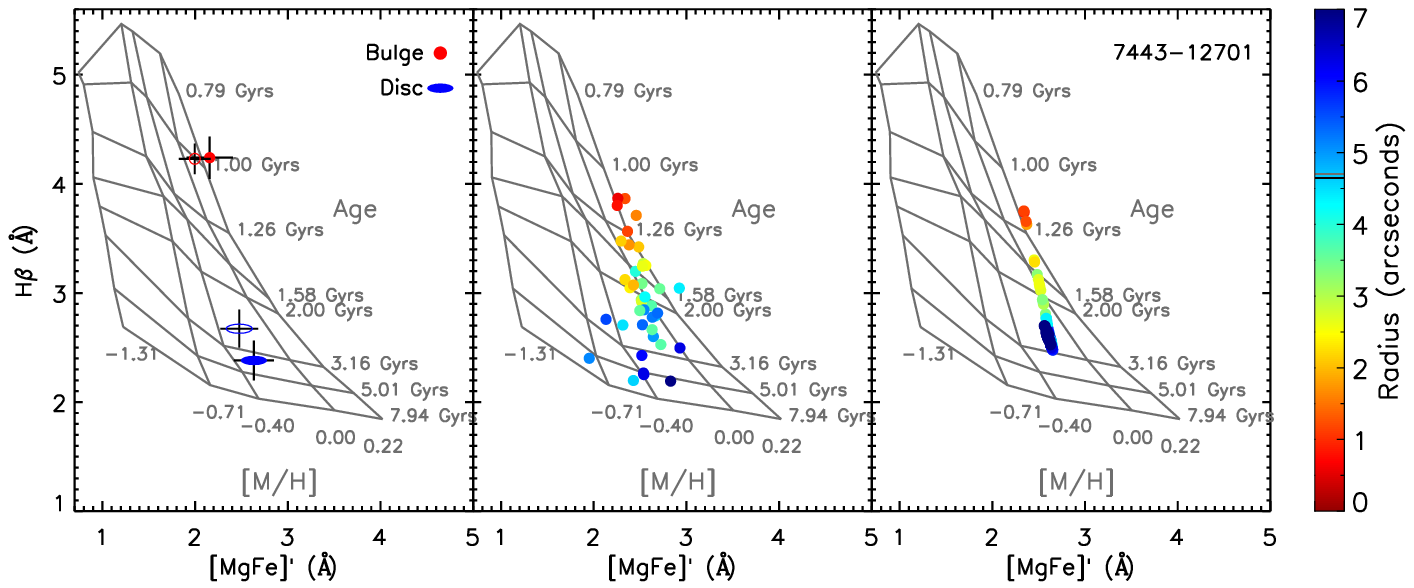}
  \includegraphics[width=1\linewidth]{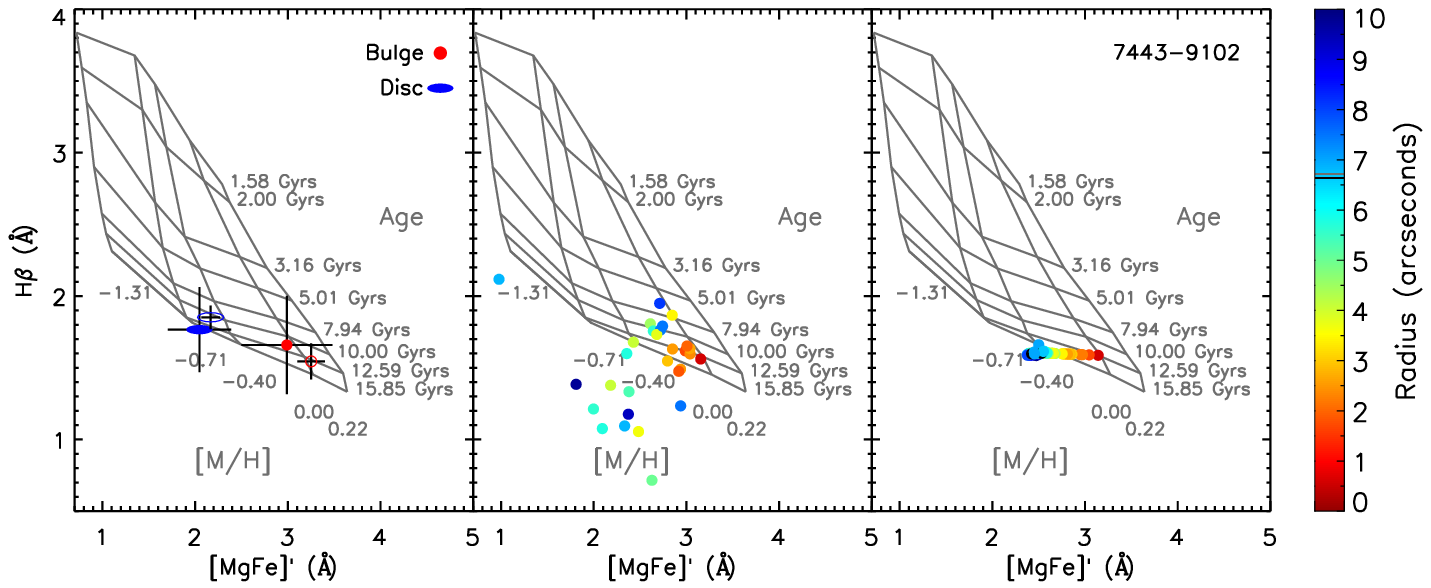}
  \caption{Index--index diagram for galaxies 7443-12701(top) and 7443-9102 (bottom) using the indices H$\beta$ versus [MgFe]$'$, with SSP model predictions of \citet{Vazdekis_2010} over plotted. From left to right are the line index measurements from the decomposed integrated bulge and disc spectra, the radially binned MaNGA datacube after obliterating the kinematics but before the decomposition, and the radially-binned bulge-plus-disc datacube obtained from the decomposition. The hollow symbols in the left plot represent the line strengths measured from the best fit spectra obtained from the regularised pPXF fits. The data points in the middle and right plots are colour-coded according to radius from the centre of the galaxy.
 \label{fig:lw_stellar_pops}}
\end{figure*}
%ssp_model_plots

The Single Stellar Population (SSP) models of \citet{Vazdekis_2010} were used to convert the line index measurements into estimates of the relative light-weighted ages and metallicities. These SSP models were created using the MILES stellar library, in which the spectra have a spectral resolution of 2.5~\AA\ (FWHM) and are convolved with a Gaussian of the appropriate dispersion to reproduce the spectral resolution of the data using a web-based tool\footnote{http://miles.iac.es/}. The resultant SSP models are thus matched to the data, minimizing the loss of information that normally occurs when degrading the data to match lower-resolution models.

An example of the one-dimensional decomposed bulge and disc spectra for galaxy 7443-12701 and the original integrated spectrum are shown in Fig.~\ref{fig:decomposed_spectra}. These spectra were obtained by fitting S\'ersic profiles to both components and identifying the bulge as the central component with the higher S\'ersic index and smaller effective radius. To further improve the fit to the image, the red star at the top  of the MaNGA field of view of that galaxy (see Fig.~\ref{fig:images}) was also included in the fit using a PSF profile to model a point source. Similarly, for 7443-9102, an additional S\'ersic profile was added to the fit to model the bright structure to the upper left of the galaxy.

The left panels of Fig.~\ref{fig:lw_stellar_pops} show the line index measurements for bulge and disc components of each galaxy plotted over the corresponding SSP model. The errors shown represent the statistical uncertainties in the line index measurements as estimated from the effect of uncertainties on the line-of-sight velocities. The global, luminosity-weighted ages and metallicities of the bulge and disc are calculated by interpolating across the SSP model grid. In 7443-12701, the bulge appears to contain younger and more metal-rich stellar populations than the disc. Since these models only give estimates of the light-weighted ages of each component, these ages shouldn't be taken as a representation of the age of the component, but more as an indication of how long ago the final episode of star formation occurred and thus when the gas was stripped out or used up. In this galaxy, the light-weighted stellar populations suggest that the final episode of star formation occurred in the bulge region. Such a result is at first surprising since this galaxy appears to be an early-type disc galaxy with no obvious spiral arms or strong emission features. However, similar trends have been noted recently in S0 galaxies. For example, \citet{Johnston_2012} and \citet{Johnston_2014} decomposed long-slit spectra of S0 galaxies in the Virgo and Fornax clusters and found evidence that the final episode of star formation in these galaxies occurred within their bulge regions. Further evidence of recent star formation in bulge regions of S0s has been detected by \citet{Poggianti_2001}, \citet{Ferrarese_2006}, \citet{SilChenko_2006b}, \citet{Kuntschner_2006} and \citet{Thomas_2006}, and a study by \citet{Rodriguez_2014} found evidence that the most recent star formation activity in discy cluster `k+a' galaxies, which are thought to be a transitional phase between spirals and S0s, was centrally concentrated within the disc. 

To check the reliability of these results, the stellar populations of the two components were compared to those from across the whole galaxy before decomposition. The kinematically obliterated datacube, obtained as an intermediate product of the decomposition outlined in Section~\ref{sec:Decomposition1}, was radially binned using the Voronoi binning technique of \citet{Cappellari_2003}, and the line indices of the binned spectra measured in the same way as the decomposed spectra. The kinematically obliterated datacube was selected for this analysis since it produces a uniform set of binned spectra with minimal variation in the line-of-sight velocities and velocity dispersions over the galaxy, and thus allows the line strengths to be over plotted on the same SSP model. These results are plotted in the top-middle SSP model of Fig.~\ref{fig:lw_stellar_pops}, colour coded according to radius after correcting for the inclination of the galaxy, and show the same trend of increasing age and decreasing metallicity with radius, albeit with more measurement noise. 

As a final comparison, this process was repeated for the decomposed bulge-plus-disc datacube. Again, the results are plotted on the top-right of Fig.~\ref{fig:lw_stellar_pops}, showing the same trend as the Voronoi-binned datacube but with reduced noise due to the smoothing in the stellar populations by the fitting process. 

It is important to remember that 7443-12701 is not a typical example of a MaNGA galaxy since it is offset from the centre of the field of view. 7443-9102 is a more typical example of a MaNGA observation for a galaxy in the Primary sample, and the same stellar populations analysis are presented in the lower panels of Figure~\ref{fig:lw_stellar_pops}. Even with the smaller field-of-view that cuts off the light from the outskirts of the galaxy, a good fit was achieved. The SSP model results for the radially-binned datacube before decomposition show a larger scatter than 7443-12701, particularly outside of 3~arcsec, which is most likely attributable to the smaller differences in the line strengths for older stellar populations. However, the estimates for the global light-weighted bulge and disc stellar populations (bottom-left panel of Fig.~\ref{fig:lw_stellar_pops}) are consistent with those the radially binned datacube before decomposition (bottom-middle panel). These consistent results provide a further indication that the decomposition technique presented in this paper can be applied to IFU data successfully, even when only the inner 1.5~R$_{e}$ of the galaxy can be observed.

As with 7443-12701, this galaxy is also likely to be an S0 galaxy, but the stellar populations show a very different trend. Due to the small differences in the H$\beta$ line strengths in older stellar populations, it can be assumed that the bulge and disc contain stellar populations of approximately similar ages. The decomposed bulge spectrum however shows a higher average metallicity than the disc spectrum. The older ages suggest that it is likely that this galaxy underwent the transformation into an S0 longer ago than 7443-12701, however the small differences in the H$\beta$ line strengths at such ages make it difficult to determine reliable age gradients across this galaxy. The metallicity gradient on the other hand suggests that the gas that fuelled the most recent star formation within the inner regions of the galaxy was more metal enriched than the gas fuelling the latest star formation in the outer regions. Since bars are thought to drive gas through the disc into the inner regions of a galaxy \citep{Kormendy_2004}, it is possible that this metallicity gradient was produced by a bar that existed before the star formation ceased.

\begin{figure*}
  \includegraphics[width=1\linewidth]{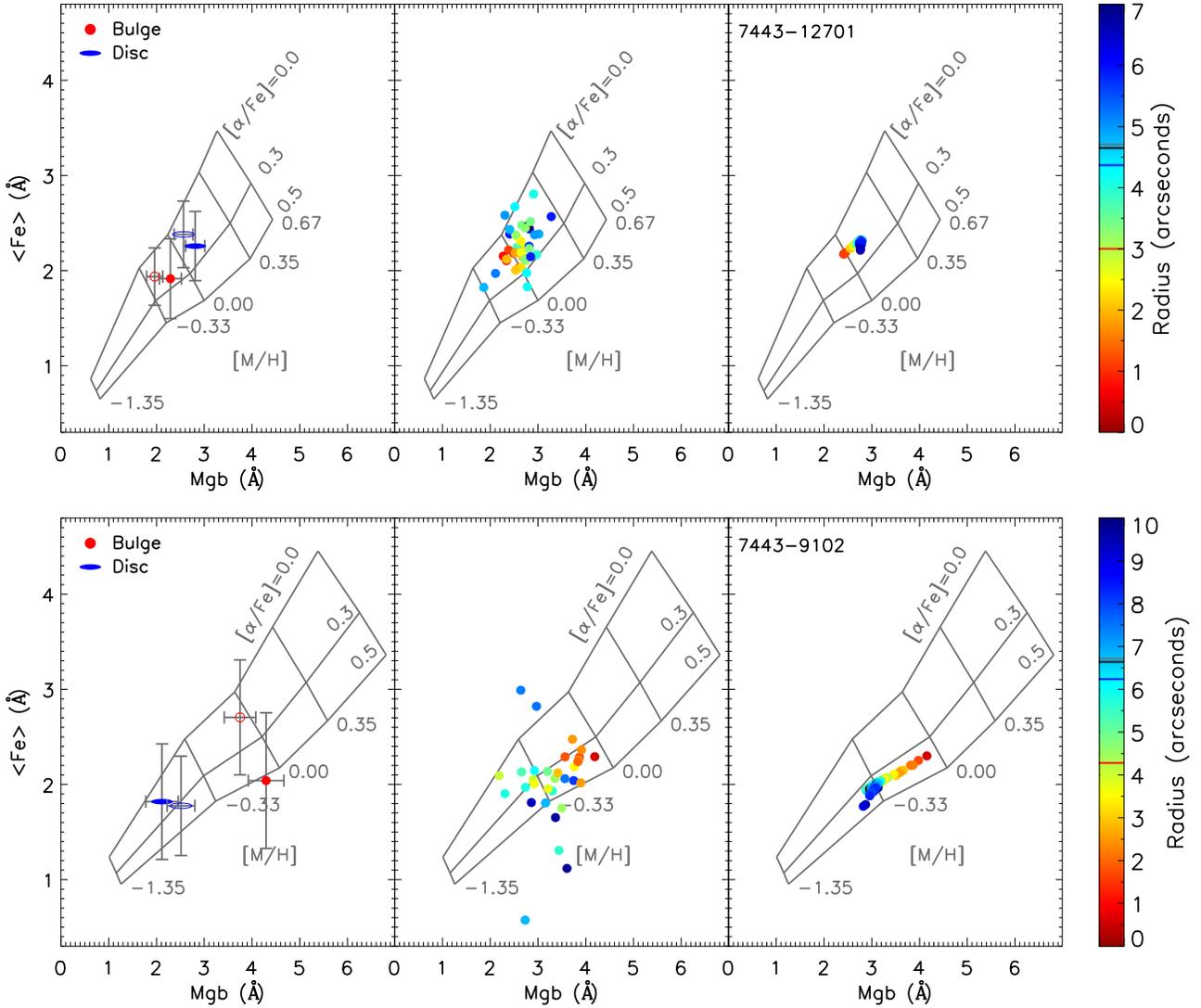}
  \caption{Index--index diagram for galaxies 7443-12701(top) and 7443-9102 (bottom) using the indices Mgb versus $\langle$Fe$\rangle$, with model predictions of \citet{Thomas_2011} over plotted. The models used for 7443-12701 and 7443-9102 are for ages of 2 and 12~Gyrs respectively, where these values are the approximate mean ages of these galaxies from Fig.~\ref{fig:lw_stellar_pops}. As in Fig.~\ref{fig:lw_stellar_pops}, the plots from left to right show the line index measurements from the decomposed integrated bulge and disc spectra, the radially binned MaNGA datacube after obliterating the kinematics but before the decomposition, and the radially-binned bulge-plus-disc datacube obtained from the decomposition. The hollow symbols in the left plot represent the line strengths measured from the best fit spectra obtained from the regularised pPXF fits. The data points in the middle and right plots are colour-coded according to radius from the centre of the galaxy.
 \label{fig:MgFe_ratios}}
\end{figure*}
%plot_MgB_Fe_paper

While the light-weighted ages and metallicities of the bulges and discs mainly tell us roughly how long ago the last episode of star formation was truncated, the magnesium-to-iron abundances can provide information on the star-formation timescales of each component. The magnesium and iron line strengths were measured, with the iron index calculated as%
\begin{equation} 
\langle\text{Fe}\rangle=(\text{Fe}5270 + \text{Fe}5335)/2.
\end{equation}
The results can be seen in Fig~\ref{fig:MgFe_ratios} for the decomposed, one-dimensional bulge and disc spectra, the radially binned datacube after obliterating the kinematics, and the radially binned bulge+disc datacube. The stellar population models of \citet{Thomas_2011} are over plotted for $\alpha$/Fe ratios of 0.0, 0.3 and 0.5, metallicities ranging between -1.35 to +0.67, and ages of 2 and 12~Gyrs, where these ages were taken as the approximate mean ages of 7443-12701 and 7443-9102 respectively from Fig.~\ref{fig:lw_stellar_pops}. 

The metallicity gradient between the bulge and disc in 7443-9102 can again be clearly seen in  Fig~\ref{fig:MgFe_ratios}. Both galaxies show super-solar $\alpha$/Fe ratios, with the bulge and disc values in each galaxy running approximately parallel to the lines of constant $\alpha$/Fe. This trend suggests that both components underwent similar star-formation timescales. A similar result was found by \citet{Johnston_2014} in their decomposed long-slit spectra of Virgo cluster S0s, and it is interesting to see the same trend appear in these two very different early-type disc galaxies. Together, these results suggest that bulge and disc formation may be tightly coupled in such galaxies, and it would be interesting to see if this conclusion remains intact with a larger sample.

The plots in Figures~\ref{fig:lw_stellar_pops} and \ref{fig:MgFe_ratios} clearly show that both the one-dimensional decomposed spectra and the two-dimensional datacube for each component form an accurate representation of the global light-weighted stellar populations of the different structural components.

\subsection{Mass-weighted stellar populations}\label{sec:MW_stellar_pops}
The light-weighted stellar populations measured within a galaxy can easily be affected by small amounts of recent star-formation, leading to the measurements being biased towards younger ages. Therefore, while these measurements are useful for determining how long ago the most recent star-formation activity occurred, they provide little information as to the star-formation history over the lifetime of the galaxy.

To get a better idea of the star-formation histories of the decomposed components, the mass-weighted ages and metallicities can instead be used. The mass-weighted ages and metallicities for each decomposed bulge and disc spectrum were calculated using the pPXF code to fit a linear combination of template spectra of known relative ages and metallicities. The MILES spectra described in Section~\ref{sec:BD_params} were used in these fits, with 300 spectra covering a regularly-sampled grid with ages in the range 0.06~-~18~Gyrs and metallicity spanning \mbox{[M/H]=~-1.71} to +0.22. To reduce the degeneracy between the ages and metallicities of the binned spectra, linear regularisation \citep{Press_1992} was used to smooth the variation in the weights of neighbouring template spectra within the age-metallicity grid. The degree of smoothness in the final solution and the quality of the final fit to the galaxy spectrum were controlled within pPXF via the regularisation parameter. While this smoothing may not completely reflect the true star-formation history of the galaxy, which is likely to be stochastic and variable over short timescales, the smoothing does reduce the age-metallicity degeneracy from bin-to-bin when applied to binned spectra from across a galaxy, thus allowing a more consistent comparison of systematic trends in the star-formation history over the whole galaxy.

The first step to measuring the mass-weighted stellar populations was to apply an unregularised fit to the bulge and disc spectra and to measure the ${\chi}^2$ value. Using this fit as the control fit, the spectrum was re-fit with a range of values for the regularisation parameter until the ${\chi}^2$ value increased by an amount equal to $\sqrt{2N}$, where $N$ is the number of pixels in the spectrum being fit. This criterion identifies the boundary between a smooth fit that still reflects the original spectrum, and a fit that has been smoothed too much and no longer represents the star-formation history of the galaxy. 

\begin{figure*}
\centering
\begin{minipage}{.5\textwidth}
  \centering
  \includegraphics[width=0.95\linewidth]{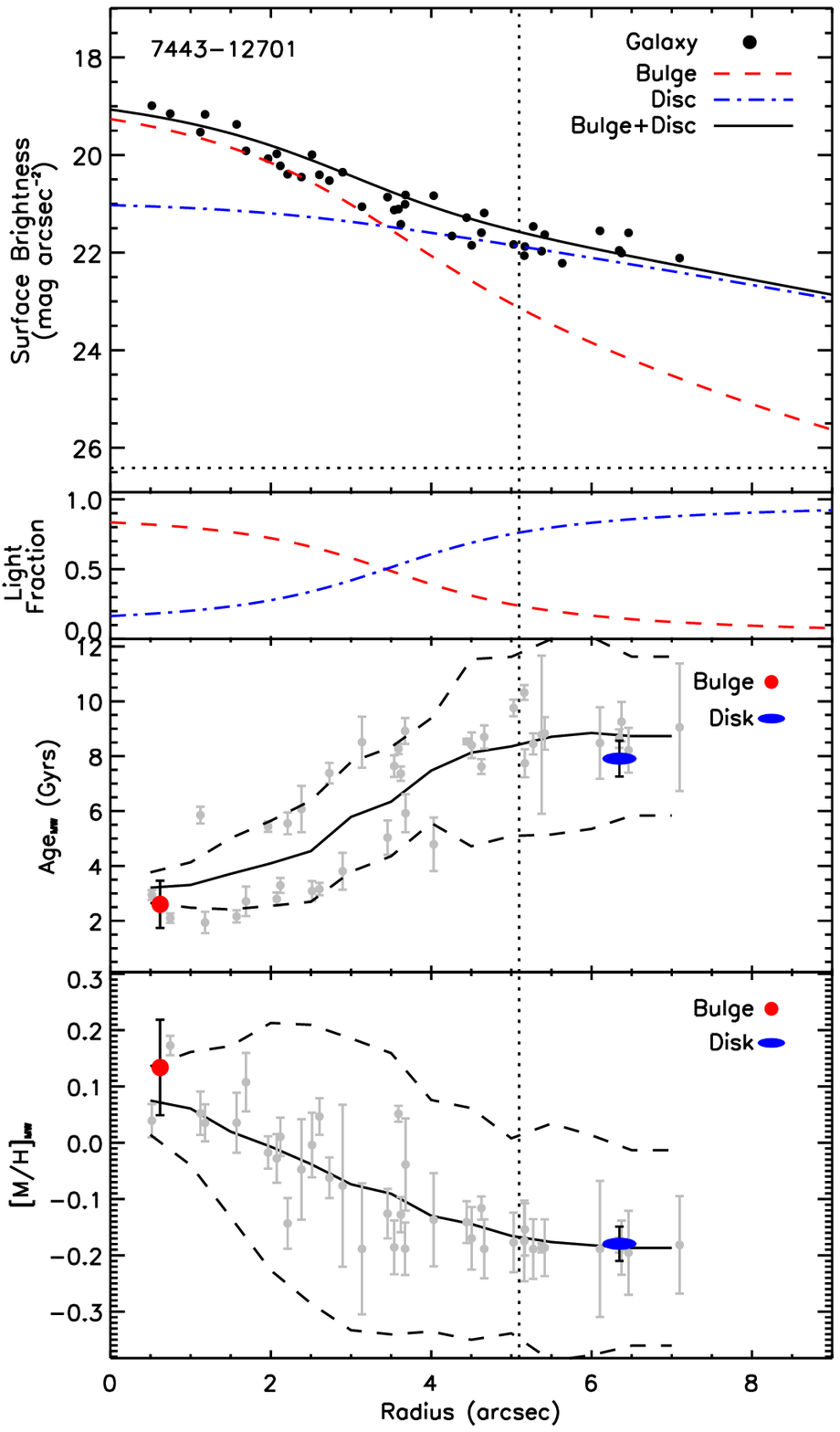}
\end{minipage}%
\begin{minipage}{.5\textwidth}
  \centering
  \includegraphics[width=0.95\linewidth]{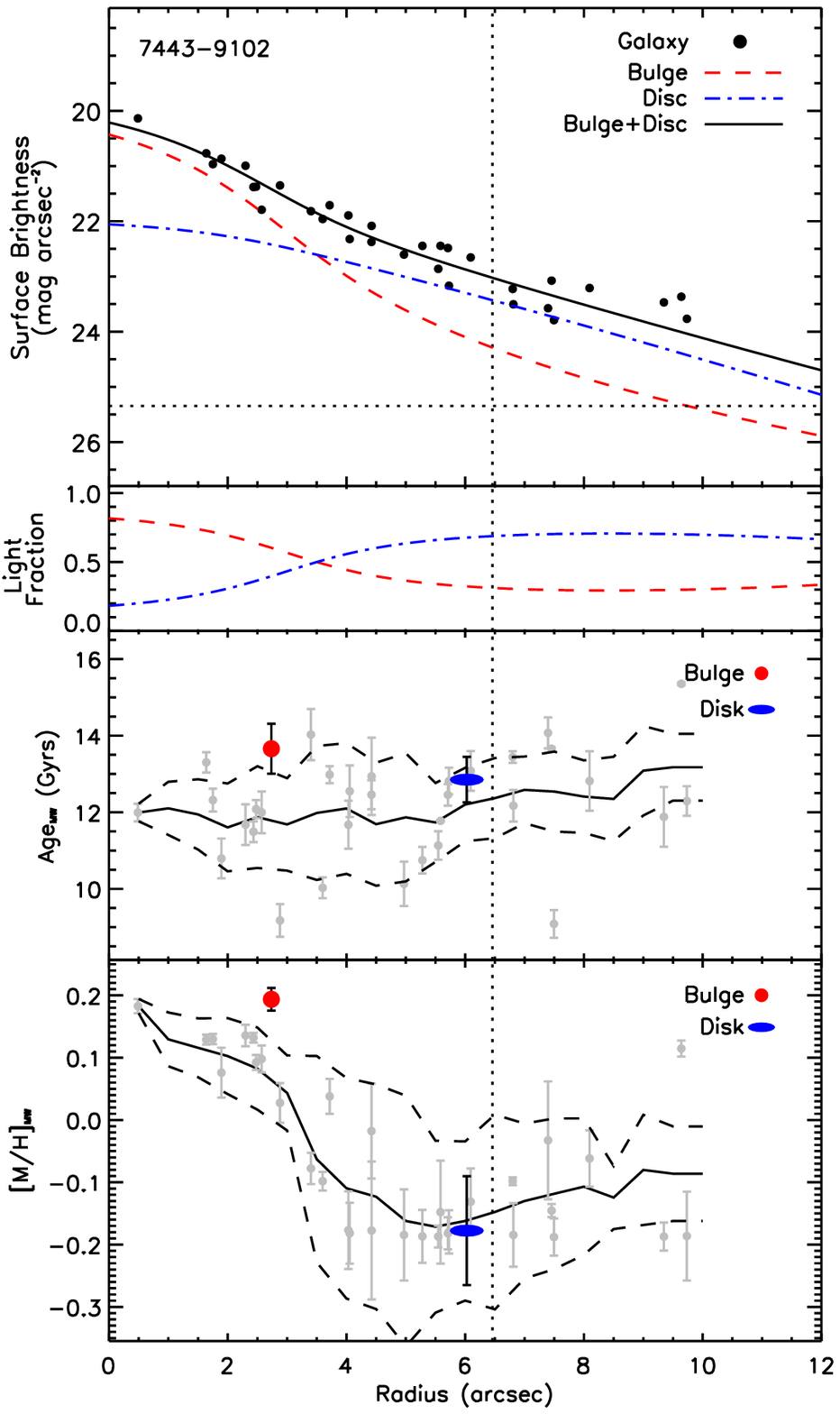}
\end{minipage}
\caption{\textit{Top}: The normalised flux for the radially-binned spectra in galaxy 7443-12701 (left) and 7443-9102 (right) against radius from the centre of the galaxy after correction for inclination. The light profile of the bulge (red dashed line) and disc (blue dot-dash line) are over plotted, as calculated from the decomposition parameters at 4770~\AA\ and convolved with a PSF of FWHM 2.5~\AA.\ The black solid line shows the bulge+disc light profile, the horizontal line gives the level of the residual-sky background from the fit, and the vertical dashed line shows the effective radius of the galaxy when fitted with a single S\'ersic profile \citep{Simard_2011} . \textit{Upper middle}: The B/T (red dashed line) and D/T (blue dot-dash line) ratios as a function of radius. \textit{Lower-middle and bottom}: The mass-weighted ages (upper) and metallicities (lower) for the bulge and disc (red circle and blue ellipse respectively), plotted against the effective radius of that component as measured from the two-component fit to the galaxy at 4770~\AA. The radially-binned measurements are shown as the grey points with their associated errors, and the solid line represents the averaged, radially-binned mass-weighted ages and metallicities. The dashed line represents the errors added in quadrature for each averaged measurement.
}
\label{fig:mw_stellar_pops}
\end{figure*}

The template spectra are modelled for an initial birth cloud mass of 1~Solar mass, and so the final weight of each template reflects the `zero-age' mass-to-light ratio of that stellar population. Therefore, the smoothed variation in the weights of neighbouring template spectra in the final regularised fit represents a simplified star-formation history of that part of the galaxy by defining the relative mass contribution of each stellar population \citep[see e.g.][]{McDermid_2015}. 
The mean mass-weighted ages and metallicities of the bulge and disc of these two galaxies were calculated using 
\begin{equation} 
	\text{log(Age$_{\text{M-W}}$)}=\frac{\sum \omega_{i} \text{log(Age$_{\text{template},i}$)}}{\sum \omega_{i}}
	\label{eq:age}
\end{equation}
and 
\begin{equation} 
	\text{[M/H]$_{\text{M-W}}$}=\frac{\sum \omega_{i} \text{[M/H]}_{\text{template},i}}{\sum \omega_{i}}
	\label{eq:met}
\end{equation}
respectively, where $\omega_{i}$ represents the weight of the $i^{th}$ template (i.e. the value by which the $i^{th}$ template stellar template is multiplied to best fit the galaxy spectrum), and [M/H]$_{\text{template},i}$ and Age$_{\text{template},i}$ are the metallicity and age of the $i^{th}$ template respectively. The results for decomposed bulges and discs of galaxies 7443-12701 and 7443-9102 are shown in Fig.~\ref{fig:mw_stellar_pops}, along with error bars estimated by Monte Carlo simulations of model galaxies constructed using the same components as in the best fit to each spectrum. These values are plotted at the effective radius of that component as measured from the two-component fit to the galaxy at 4770~\AA\ within the datacube. This wavelength is the central wavelength of the \textit{g}-band, whose bandwidth covers the H$\beta$, magnesium and iron lines used in Section~\ref{sec:LW_stellar_pops}. Although the estimates for the light- and mass-weighted stellar populations of each component are different, the same trends can be seen. 7443-12701 for example still shows the younger and more metal rich bulge compared to its disc, while in 7443-9102 the bulge and disc ages are similar while the bulge shows higher metallicity than the disc. As a further check on the reliability of these results, the line strengths of the best fit to each decomposed spectrum were measured and compared with those of the decomposed spectra. The line strengths were found to be consistent between the decomposed and best fit spectra, as can be seen in Figures~\ref{fig:lw_stellar_pops} and \ref{fig:MgFe_ratios}.

%% ifu_stellar_pops_radial.pro

As in Fig.~\ref{fig:lw_stellar_pops}, the mass-weighted ages and metallicities of the Voronoi-binned spectra are also plotted against the radius of each bin, having been measured in the same way.  The radially averaged mass-weighted ages and metallicities are overplotted as the solid line, and the uncertainty in this trend marked as the dashed lines as measured by adding in quadrature the errors on the measurements used in the radial binning. In both cases, the radial trend follows the trend between the bulge and disc, indicating that no significant information was lost from the bulge and disc spectra as a result of the decomposition process. The scatter is simply due to the variations in the stellar populations over the galaxy. 

For comparison, Fig.~\ref{fig:mw_stellar_pops} also shows the light profile of each galaxy, as measured from the broad-band image covering the centre of the \textit{g}-band (4770~\AA), along with the S\'ersic profiles for the bulge and disc and the bulge-plus-disc profile at that wavelength. These profiles have been convolved with the appropriate PSF to reflect the spatial resolution of the final datacube. The level of the residual-sky background, as measured by \textsc{GalfitM}, and the effective radius measured from a single S\'ersic fit \citep[\textit{g}-band, ][]{Simard_2011} are also plotted at the horizontal and vertical dotted lines as a reference. This plot, combined with the plot for the variation in the B/T and Disc-to-Total (D/T) light ratios with radius, indicates the proportion of bulge and disc light at each radius in the SSP models, and thus the radius at which the disc dominates the light of the galaxy.

\begin{figure}
  \includegraphics[width=1\linewidth]{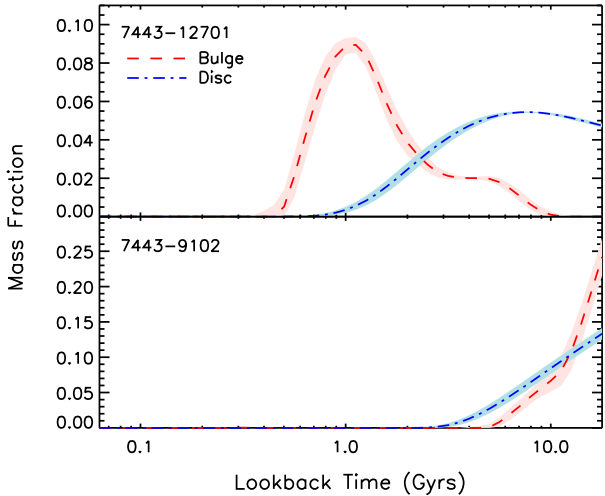}
  \caption{The star-formation histories of the bulges (red) and discs (blue) of galaxies 7443-12701 (top) and 7443-9102 (bottom). The star-formation history is defined as the relative fraction of stellar mass created as a function of the lookback time, and was derived from the regularised fit to the bulge and disc spectra with MILES template spectra. The shaded region around each line shows the rolling standard error on the mean value at each lookback time, as estimated from the standard deviation in the mass fractions in each age bin divided by the square root of the number of measurements used in the bin.
\label{fig:mass_fractions}}
\end{figure}

The regularised weights of each template from PPXF reflects the relative mass contribution of that stellar population, and therefore the mass created at that time. Using these weights, the estimated star-formation histories of the bulges and discs of galaxies 7443-12701 and 7443-9102 were derived, and are shown in Figure~\ref{fig:mass_fractions}. The star-formation histories are plotted as the mass fraction of each component created as a function of time, where the smooth variation is a result of the regularisation outlined above. The standard error on the mass fractions are marked as the shaded regions, estimated from the standard deviation in the mass fractions in each rolling-age bin divided by the square root of the number of measurements used in the bin. In 7443-9102, the star formation in the bulge and disc ceased a long time ago, reflecting the results from the light-weighted stellar populations analysis. 7443-12701 on the other hand shows more recent and extended star formation in both components. Within the disc, the star formation has slowly truncated since $\sim$8~Gyrs ago, whereas two star-formation events can be seen in the bulge at $\sim$5 and $\sim$1~Gyrs ago. This recent episode of star formation in the bulge dominates the luminosity of the bulge, and thus biases the light-weighted age of the bulge to younger values.

%mass_fractions.pro

\subsection{Decomposition and analysis of simulated datacubes}\label{sec:decomposed_datacubes}
 
 The example analyses presented above show promising results that the decomposed bulge and disc spectra reliably reflect the stellar populations and star-formation histories of those components. However, as a final test of the reliability of the the technique presented in this paper, the decomposition and analysis was repeated for a series of simulated galaxies created using the decomposed bulge and disc spectra of 7443-12701 and 7443-9102.

\begin{figure}
  \includegraphics[width=1\linewidth,trim={0 2cm 0 0},clip]{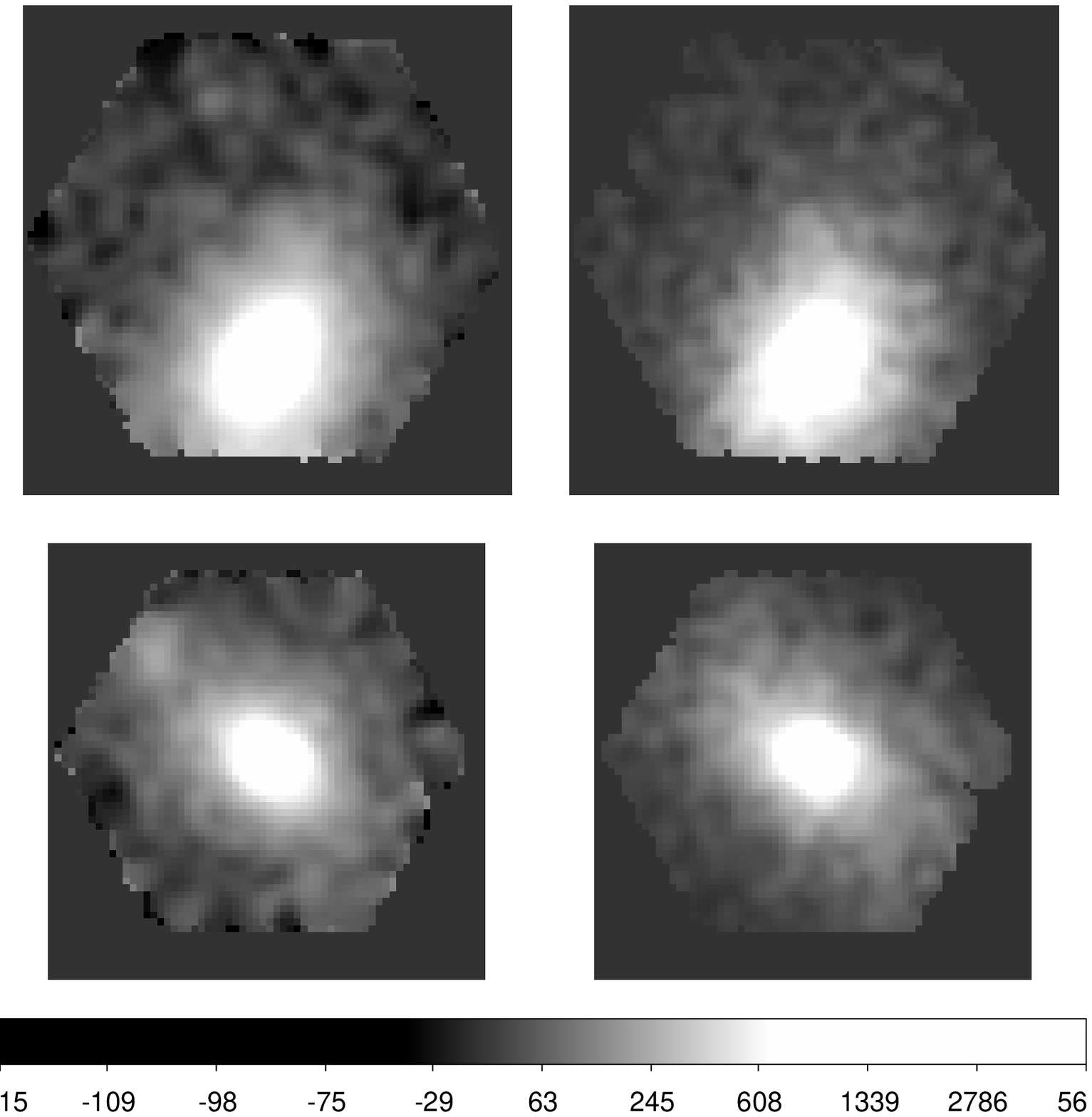}
  \caption{A comparison of image slices from the original (left) and simulated (right) datacubes for 7443-12701 and 7443-9102 (top and bottom respectively) at 4990~\AA. The images have been scaled logarithmically on the same flux scale, which has been adjusted to maximally display the small-scale variations in the background.
\label{fig:simulations_images}}
\end{figure}

For each galaxy, 10 simulated galaxy datacubes were created by combining two datacubes representing the galaxy and the background noise. The galaxy datacube was created by adding together the decomposed bulge and disc datacubes with the residual sky background. To achieve a realistic test, a sensible amount of noise must be added to this data cube,  
which could technically be done by simply adding poisson noise to each pixel in this cube. However, as
MaNGA datacubes are created by combining dithered pointings with each fibre-IFU, the level of the noise in neighbouring spaxels is correlated in those datacubes, and therefore must be recreated in the simulated noise. An example of this effect can be seen in the images on the left of Fig.~\ref{fig:simulations_images}, in which the flux scales have been adjusted to display the small-scale variations in the background noise. 
This noise correlation must be reflected in the noise added for this simulated data cube. To achieve this effect, the noise data cubes - before adding them onto the noise-less data cube  - were convolved with a PSF profile of varying FWHM until the structure of the noise resembled that within the MaNGA cubes.
Finally, the simulated galaxy datacubes were combined with the noise datacubes and compared by eye to the original datacubes at the same wavelength. When using the Poissonian noise levels alone as the starting point for the noise datacube, due to the smoothing the final simulated datacube was found to have too low levels of background variations when compared to the original datacube at the same flux level. Instead, the closest similarity was achieved when the Poisson noise was multiplied by a factor of 5 and convolved with a PSF of 3 pixels FWHM for both galaxies. Figure~\ref{fig:simulations_images} gives a comparison of the original and simulated datacubes at 4990~\AA\ on the same flux scales.

The simulated datacubes were decomposed in the same way as the original MaNGA datacubes, and the analysis described in Sections~\ref{sec:LW_stellar_pops} and \ref{sec:MW_stellar_pops} applied to the decomposed spectra. At this point it is important to note that, because we do not smooth the noise in the simulated datacubes in the spectral direction, the decomposed simulated spectra contain higher levels of noise than the original decomposed spectra since they additionally include the noise that was already present in the decomposed bulge and disc datacubes. The analysis here hence presents a worst-case scenario, which likely over-predicts the error bars of our analysis.

\begin{figure}
  \includegraphics[width=1\linewidth]{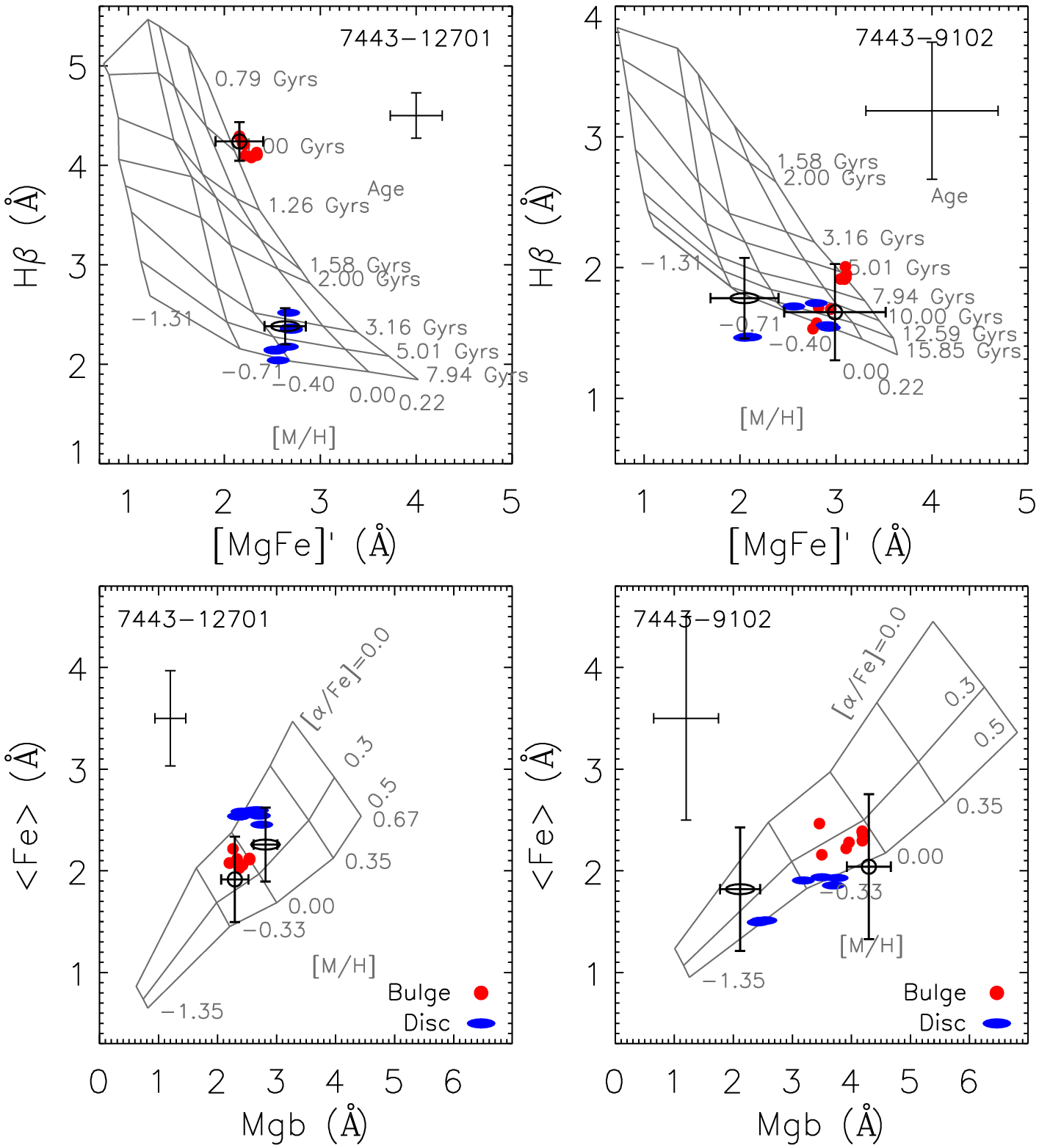}
  \caption{Index-index diagrams for the decomposed bulge and disc spectra from each simulated galaxy based on 7443-12701 (left) and 7443-9102 (right). As in Figures~\ref{fig:lw_stellar_pops} and \ref{fig:MgFe_ratios}, the models over plotted in the top row are those of \citet{Vazdekis_2010}, and the models of \citet{Thomas_2011} are used in the bottom row. The hollow black circle and ellipse represent the results for the decomposed bulge and disc spectra respectively from the original datacube. The mean uncertainties for each plot are shown in the top right or top left of each plot.
\label{fig:simulations_comparisons_lw}}
\end{figure}

The light-weighted stellar populations analysis for each simulated galaxy are shown in Fig.~\ref{fig:simulations_comparisons_lw}, and appear consistent with those in Figures~\ref{fig:lw_stellar_pops} and \ref{fig:MgFe_ratios} (as shown by the hollow symbols). The larger uncertainties and the scatter in the simulated galaxy results were found to simply reflect the higher noise levels in those datacubes, which resulted in noisier decomposed spectra.
The largest discrepancy between the real and simulated results are in the values for the line indices measured in the decomposed discs of the simulated galaxies based on 7443-9102, which is most noticeable in the bottom right plot of Fig.~\ref{fig:simulations_comparisons_lw}. While the results from 6 of the 10 simulated discs agree very well with each other and lie within the error bars of the original decomposed disc measurements, the remaining four measurements show larger scatter. When looking at the line index measurements across the decomposed bulges and discs of each simulation, it was found that if one component overestimated the line strength, the other underestimated it proportionally. This effect likely arose from the lower S/N in the simulations in comparison to the original datacubes, which led to larger variation in the magnitudes of the fits. Since the decomposed bulge and disc spectra of galaxy 7443-9102 appears noisier than those of 7443-12701, and consequently the corresponding simulations also contain higher levels of noise, this effect is stronger in this case. Furthermore, it appears strongest in the Mgb index, potentially because it forms part of a triplet which is blurred together in the first step of the decomposition process outlined in Section~\ref{sec:Decomposition}.

\begin{figure}
  \includegraphics[width=1\linewidth]{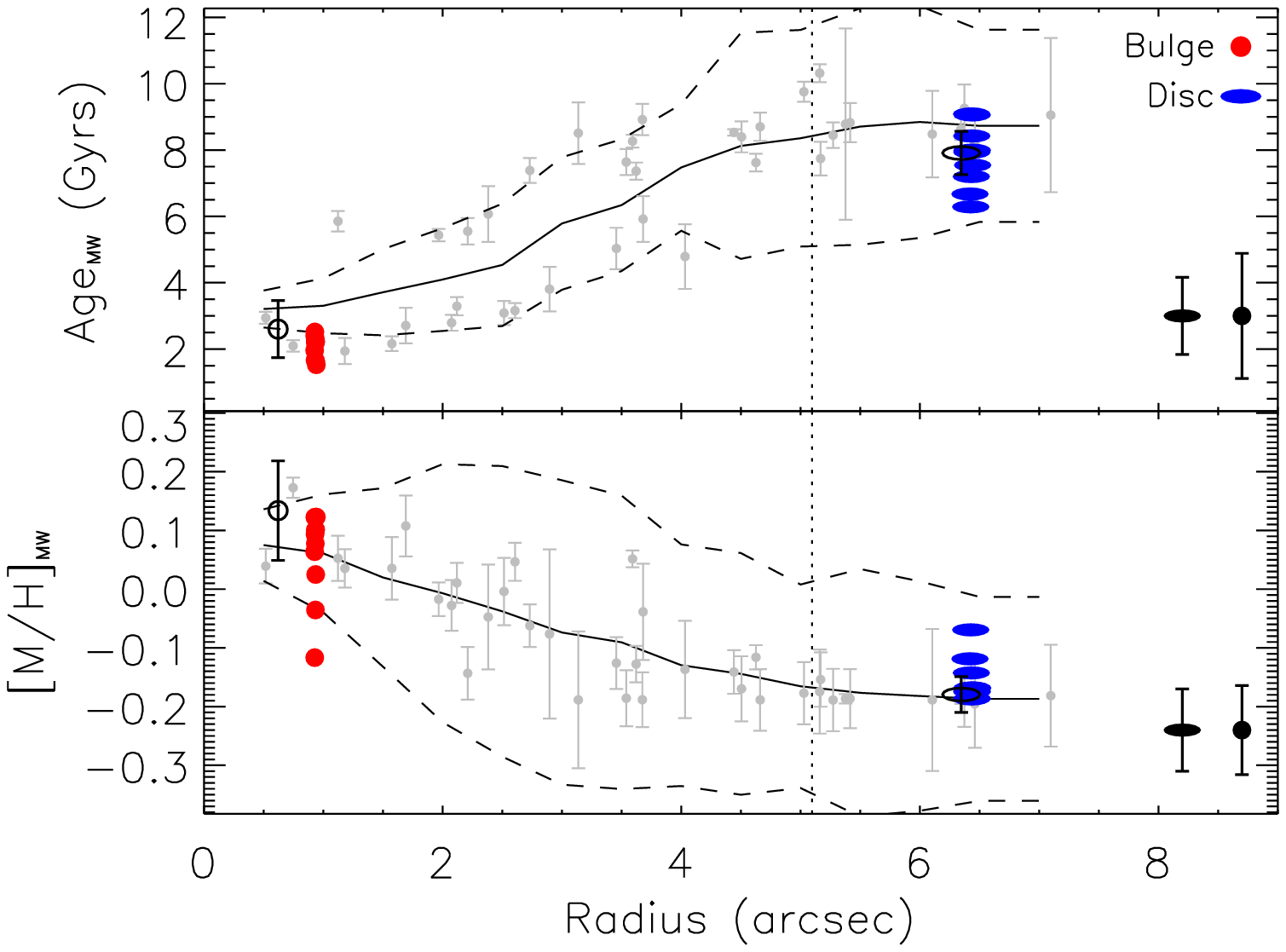}
  \includegraphics[width=1\linewidth]{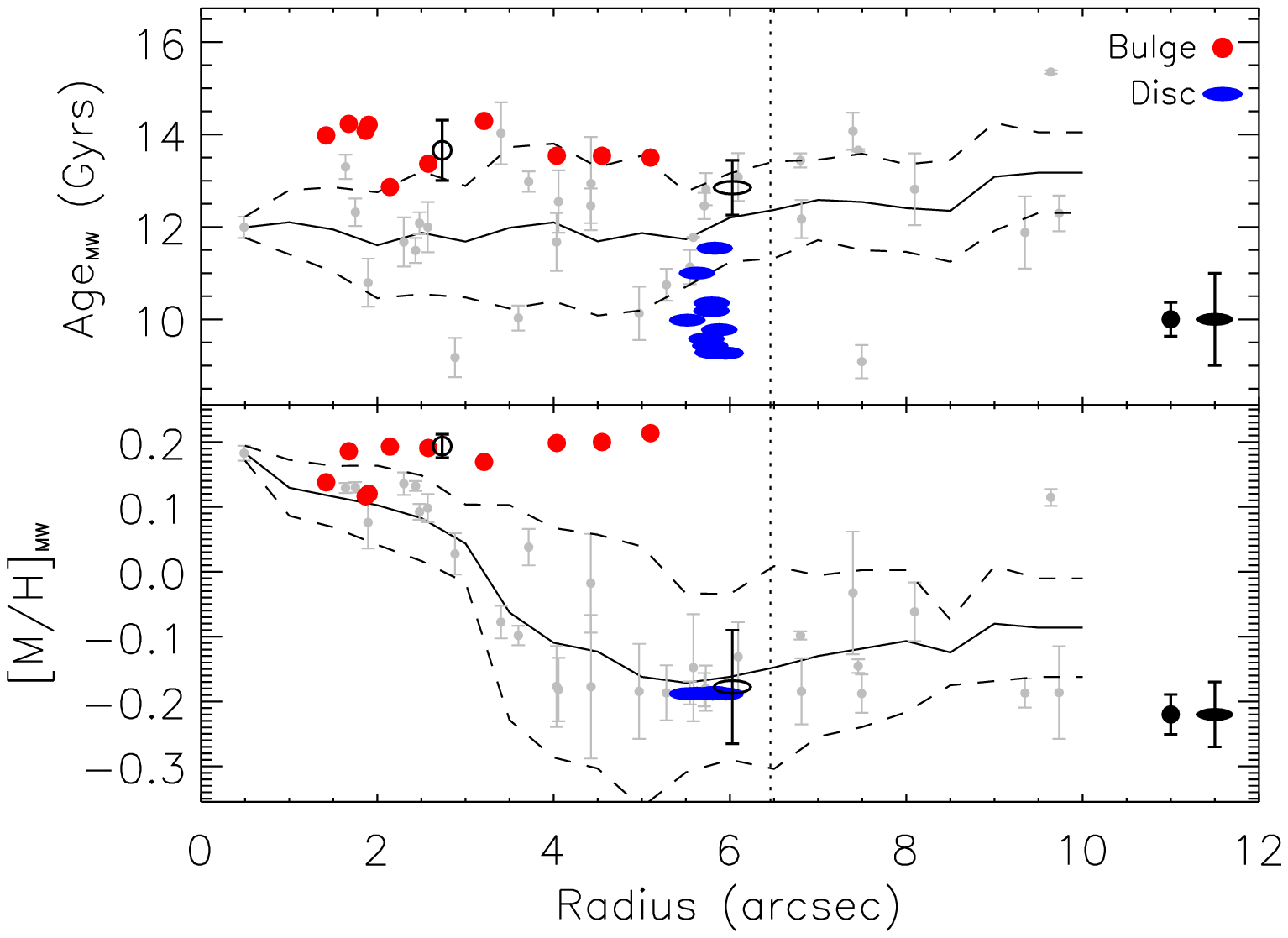}
  \caption{The mass-weighted ages (upper) and metallicities (lower) for the bulges and discs (red circle and blue ellipse respectively) of the simulated galaxies, plotted against the effective radius of that component as measured from the two-component fit to the galaxy at 4770~\AA. As in Fig~\ref{fig:mw_stellar_pops}, the radially-binned measurements from the original datacubes are shown as the grey points with their associated errors, along with the averaged, radially-binned mass-weighted ages and metallicities and the errors added in quadrature for each averaged measurement. \textsc{red}{As a reference, }the black circles and ellipses represent the results for the decomposed bulge and disc spectra respectively from the original datacube, \textsc{red}{as shown in Fig.~\ref{fig:MgFe_ratios}}. The mean uncertainties in the simulated bulge and disc measurements are shown in the bottom right corner of each plot.
   \label{fig:simulations_comparisons_mw}}
\end{figure}

The mass-weighted stellar populations for the simulated galaxies are presented in Fig.~\ref{fig:simulations_comparisons_mw}, overplotted with the results from the radially binned and decomposed spectra from Fig.~\ref{fig:mw_stellar_pops} as a comparison. Each age and metallicity measurement is plotted at the effective radius of that component, as measured during the decomposition of that simulated galaxy. Again the results appear largely consistent with those obtained from the original galaxies, with the main exception being the ages of the simulated discs in 7443-9102. This discrepancy may be due to the higher levels of noise in the decomposed spectra as well as the small variations in the line strengths at such high ages. It is interesting to note that the effective radius of the bulge of the simulated galaxies based on 7443-9102 also show a large variation. The original fit for this galaxy contained a high-S\'ersic index bulge, which consequently was used in the simulated galaxies. It is likely that the variation in the measured effective radii of the bulges of the simulated galaxies is simply due to the combination of the high S\'ersic index, small size and elevated noise levels in the simulated galaxies. However, despite the variation in their sizes, the mass-weighted ages and metallicities appear consistent across the 10 simulated galaxies. This result is consistent with the finding of \citet{Johnston_2014}, who found that the stellar populations derived from decomposed long-slit spectra were fairly robust against uncertainties in the decomposition parameters used in the fits.

\begin{figure}
  \includegraphics[width=1\linewidth]{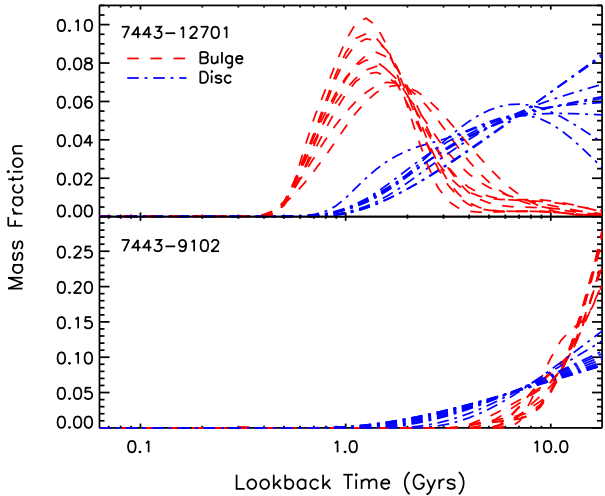}
  \caption{The star-formation histories of the bulges (red) and discs (blue) of the simulated galaxies based on 7443-12701 (top) and 7443-9102 (bottom). The star-formation history is defined as the relative fraction of stellar mass created as a function of the lookback time, and was derived from the regularised fit to the bulge and disc spectra with MILES template spectra. 
\label{fig:mass_fractions_simulations}}
\end{figure}

Finally, Fig.~\ref{fig:mass_fractions_simulations} shows the derived star-formation histories for each simulated galaxy. While these plots do show some scatter, especially in the case of 7443-12701, the general trends are consistent between each model and with those presented in Fig.~\ref{fig:mass_fractions}. The lower S/N of the decomposed spectra derived from the simulated galaxies is the most likely reason for this scatter. The systematically lower ages of the simulated discs in star-formation history of the discs in Fig~\ref{fig:simulations_comparisons_mw} are also reflected in the star-formation histories of the simulated discs by extending to younger ages.

These finals tests show that where the model gives a good fit to the original galaxy, the decomposed spectra are a good representation of the star-formation histories of each component within the galaxy. However, care must be taken when decomposing datacubes of low S/N since the noise can propagate through to the final decomposed spectra, as shown here.

%plot_simulation_results.pro

%ifu_stellar_pops_radial

%mass_fractions_tests2.pro
\smallskip
\smallskip
\smallskip

\section{Discussion}\label{sec:Summary}

We have presented \textsc{buddi}, a new method for applying two-dimensional bulge--disc decomposition to IFU data to cleanly separate the light from the bulges and discs in order to study their independent star-formation histories. The decomposed spectra can be displayed as a datacube for each component, or as an integrated, one-dimensional spectrum. This technique builds upon earlier work, which decomposed multi-waveband photometry or long-slit spectra, to combine both the spectral and spatial information in the decomposition. As a result, a cleaner separation of the bulge and disc stellar populations can be achieved, leading to a better understanding of their independent star-formation histories and thus how galaxies evolve.

Since IFU datacubes typically have a smaller field of view than imaging data, a series of tests have been carried out on both simulated and real MaNGA data in order to determine the reliability of the decomposition when the outskirts of the galaxy may not be visible. These tests found that in cases where the galaxy can be well modelled by a single or double S\'ersic model, the best fit using the MaNGA field of view for the 127, 91 and 61-fibre IFUs is comparable to that using the larger field of view typically available for imaging data. This result suggests that by using the full spectral information through the datacube and the increased signal-to-noise over the whole dataset when applying simultaneous fits, the fit to each galaxy is reliable and can accurately separate the spectra from each component. Additional tests were carried out to compare the single-band fits from \textsc{Galfit} with the simultaneous fitting of \textsc{GalfitM}. These tests found that \textsc{GalfitM} improved the extraction of information, particularly at low S/N, which is consistent with the comparisons carried out by \citet{Vika_2013} on a smaller number of wavebands.

In order to demonstrate the capabilities and applications of this new technique, a preliminary stellar populations analysis is presented for two early-type galaxies from the MaNGA commissioning sample. The light- and mass-weighted stellar populations were measured for both the decomposed bulge and disc spectra and for the radially binned datacubes after obliterating the kinematics but before decomposition. The results from the decomposed spectra were found to be less noisy than and consistent with those from the radially-binned datacubes, again showing that the decomposition of this type of data is reliable for a range of galaxy morphologies and thus can be useful for detailed studies of galaxy evolution through the build-up of the different components. Similarly, the results derived from decomposing a series of simulated datacubes created from the decomposed spectra for 7443-12701 and 7443-9102 were found to be reproducible and consistent with the measurements from these galaxies, further reflecting the reliability of the technique when an appropriate model is selected for the fit.

The tests and analysis presented in this paper already illustrate the potential power of applying profile-fitting techniques to IFU data, but there is still room for development. For a more complete analysis, the technique could be applied to other datasets such as CALIFA or MUSE data. Additionally, one could use the residual datacubes, created by subtracting the best fit from the original data, to analyse the stellar populations in non-smooth structures such as spiral arms. The same analysis could also be carried out using the new non-parametric feature within \textsc{GalfitM}, which is still being tested by the MegaMorph team.

It would also be useful to make better use of the kinematics information that is currently obliterated in this technique. For example, any emission features present could be modelled and subtracted to allow for a better measurement of the stellar populations in currently star-forming galaxies, thus extending this technique to a wider range of galaxy types. 

While the analysis presented in this paper is able to measure colour gradients within each component through the variation in the effective radius of that component with radius \citep{Johnston_2012, Vulcani_2014, Kennedy_2015}, the degeneracy between age and metallicity prevent this information from being translated into stellar population gradients. However, in \citet{Johnston_2012} we showed that by allowing the effective radii to be free with wavelength in the fits of long-slit spectra, the line index gradients could be measured and used to derive gradients in the ages and metallicities. This idea could be applied to the analysis above by repeating the fits over smaller wavelength regions centred on the spectral feature of interest and reducing the constraints on the effective radii, which could then be translated into age and metallicity gradients within each component. Such an improvement in the technique described here could significantly increase the amount of information we could learn about the independent star-formation histories of each component, and the cost of larger error bars due to using less data.

By isolating the bulge and disc stellar populations and the gas and stellar kinematics in this way, we plan to address open questions about their evolution. For example, one could go beyond deriving the star-formation histories of the bulge and disc as a whole, and instead analyse how the star formation was triggered and truncated in the inner and outer regions of each component. Positive age gradients with radius within the discs of S0s could reflect a gas stripping scenario leading to the truncation of star formation from the outside in \citep{Larson_1980, Bekki_2002}, while flatter gradients may correspond to interactions with neighbouring galaxies leading to a final episode of star formation throughout the disc \citep{Mihos_1994}. Similarly, negative age gradients within bulges could indicate that the bulge was built up through multiple minor mergers \citep{Kormendy_2004}, while positive age gradients would reflect a scenario where gas was funnelled into the centre where it fuelled more recent star formation \citep{Friedli_1995}. 

All these ideas, however, are well beyond the scope of this paper and will be addressed in future publications.

%\newpage
\section*{Acknowledgements}

We would like to thank Vladimir Avila-Reese for useful discussions about the technique and future applications of the technique presented in this paper.    We would also like to thank the anonymous referee for their useful comments that helped improve this paper.

EJJ acknowledges support from the Marie Curie Actions of the European Commission (FP7-COFUND) and the Science and Technology Facilities Council (STFC). KB acknowledges support from the World Premier International Research Center Initiative (WPI Initiative), MEXT, Japan and by JSPS KAKENHI Grant Number 15K17603. MAB acknowledges support from NSF AST-1517006.

Funding for the Sloan Digital Sky Survey IV has been provided by
the Alfred P. Sloan Foundation, the U.S. Department of Energy Office of
Science, and the Participating Institutions. SDSS-IV acknowledges
support and resources from the Center for High-Performance Computing at
the University of Utah. The SDSS web site is www.sdss.org.

SDSS-IV is managed by the Astrophysical Research Consortium for the 
Participating Institutions of the SDSS Collaboration including the 
Brazilian Participation Group, the Carnegie Institution for Science, 
Carnegie Mellon University, the Chilean Participation Group, the French Participation Group, Harvard-Smithsonian Center for Astrophysics, 
Instituto de Astrof\'isica de Canarias, The Johns Hopkins University, 
Kavli Institute for the Physics and Mathematics of the Universe (IPMU) / 
University of Tokyo, Lawrence Berkeley National Laboratory, 
Leibniz Institut f\"ur Astrophysik Potsdam (AIP),  
Max-Planck-Institut f\"ur Astronomie (MPIA Heidelberg), 
Max-Planck-Institut f\"ur Astrophysik (MPA Garching), 
Max-Planck-Institut f\"ur Extraterrestrische Physik (MPE), 
National Astronomical Observatory of China, New Mexico State University, 
New York University, University of Notre Dame, 
Observat\'ario Nacional / MCTI, The Ohio State University, 
Pennsylvania State University, Shanghai Astronomical Observatory, 
United Kingdom Participation Group,
Universidad Nacional Aut\'onoma de M\'exico, University of Arizona, 
University of Colorado Boulder, University of Oxford, University of Portsmouth, 
University of Utah, University of Virginia, University of Washington, University of Wisconsin, 
Vanderbilt University, and Yale University.

This research made use of Montage. It is funded by the National Science Foundation under Grant Number ACI-1440620, and was previously funded by the National Aeronautics and Space Administration's Earth Science Technology Office, Computation Technologies Project, under Cooperative Agreement Number NCC5-626 between NASA and the California Institute of Technology.

\footnotesize{
\bibliographystyle{mn2e}

\bibliography{paper4_refs}
}

\end{document}